\documentclass[aps,10pt,pra,twocolumn,groupedaddress,showpacs,showkeys,amsfonts]{revtex4-1}
\usepackage{amsfonts}
\usepackage{amsmath}
\usepackage{amssymb}
\usepackage{graphics,graphicx}
\graphicspath{{./figures/}}
\usepackage{epsf}
\usepackage{subfigure}
\usepackage{hyperref}
\usepackage[all]{hypcap}
\usepackage{color}
\newcommand{\eq}[1]{Eq.~\eqref{#1}}
\newcommand{\eqs}[1]{Eqs.~\eqref{#1}}
\newcommand{\eqss}[2]{Eqs.~\eqref{#1}-\eqref{#2}}
\newcommand{\seq}[1]{Sec.~\ref{#1}}
\newcommand{\app}[1]{App.~\ref{#1}}
\newcommand{\fig}[1]{Fig.~\ref{#1}}

\newcommand{\be}{\begin{equation}}
\newcommand{\ee}{\end{equation}}
\newcommand{\bem}{\begin{multline}}
\newcommand{\bea}{\begin{align}}
\newcommand{\eea}{\end{align}}
\def\mean#1{\left< #1 \right>}
\def\cro#1{\left[ #1 \right]}
\def\pare#1{\left( #1 \right)}

\def\abs#1{\lvert{ #1 \rvert}}

\begin{document}

\title{Far-from-equilibrium noise heating and laser cooling dynamics in radio-frequency Paul traps}

\author{A. Maitra$^1$}
\author{D. Leibfried$^2$}
\author{D. Ullmo$^1$}
\author{H. Landa$^{3}$}
\email{haggaila@gmail.com}
\affiliation{$^1$LPTMS, CNRS, Univ.~Paris-Sud, Universit\'e Paris-Saclay, 91405 Orsay, France
\\$^2$National Institute of Standards and Technology, 325 Broadway, Boulder, Colorado 80305, USA \\ $^3$Institut de Physique Th\'{e}orique, Universit\'{e} Paris-Saclay, CEA, CNRS, 91191 Gif-sur-Yvette, France}

\begin{abstract}

We study the stochastic dynamics of a particle in a periodically driven potential. For atomic ions trapped in radio-frequency Paul traps, noise heating and laser cooling typically act slowly in comparison with the unperturbed motion. These stochastic processes can be accounted for in terms of a probability distribution defined over the action variables, which would otherwise be conserved within the regular regions of the Hamiltonian phase space. We present a semiclassical theory of low-saturation laser cooling applicable from the limit of low-amplitude motion to large-amplitude motion, accounting fully for the time-dependent and anharmonic trap. We employ our approach to a detailed study of the stochastic dynamics of a single ion, drawing general conclusions regarding the nonequilibrium dynamics of laser-cooled trapped ions. We predict a regime of anharmonic motion in which laser cooling becomes diffusive (i.e., it is equally likely to cool the ion as it is to heat it), and can also turn into effective heating. This implies that a high-energy ion could be easily lost from the trap despite being laser cooled; however, we find that this loss can be counteracted using a laser detuning much larger than Doppler detuning.

\end{abstract}


\maketitle

\section{Introduction}\label{Sec:Intro}

In the past 50 years, Paul traps have become a major tool for confining charged particles \cite{paul1990electromagnetic}. Certain atomic ion species can be cooled with laser light over more than six orders of magnitude in temperature to the millikelvin regime \cite{wineland1979laser}, and with suitable methods further to their quantum ground state, making Paul traps a prominent tool in experiments demonstrating quantum control \cite{RevModPhys.85.1103,PhysRevLett.117.060504}. However, the nonequilibrium dynamics leading ions from the high energies at creation or after a collision with background gas to the near-equilibrium ``Doppler cooling limit'' remain poorly understood. Partly, the reason is that Paul traps are based on radio-frequency (rf) electric fields whose interplay with additional
external fields that act on the ions poses a theoretical and experimental challenge.

{The rapidly oscillating fields of the Paul trap lead to an averaged effective trapping potential with (slower) motion at the characteristic ``secular'' frequencies. Superimposed on the secular motion is a smaller-amplitude motion at the rf-drive frequency called ``micromotion''. Additional, periodic rf-driven motion that is independent of the secular motion amplitude, known as ``excess micromotion'', can often be carefully reduced, and we assume here that it can be neglected. 
The interplay of rf-driven motion and stochastic noise forms a basic example of a system driven far from equilibrium, and has been a concern since the early experiments with trapped ions.} The potential in the early Paul traps was typically quadrupolar near the effective potential minimum, leading to linear equations of motion. In \cite{BrownianModel} a model for a one-dimensional (1D) periodically driven Mathieu oscillator subject to dissipation and white noise generalized previous works on Brownian motion. The Fokker-Planck (FP) equation for the probability distribution of the particle in phase-space (of position and momentum) admits a Gaussian solution that is periodic with the trap frequency \cite{LangevinIon,alekseev1994effect, sobvehart1994singular}. 
The interplay of the periodic driving with laser cooling was elaborated for a quadrupole trap in some parameter regimes \cite{cirac1994laser}, mostly in the final stage of the cooling when the ion settles around the effective potential minimum, and also with excess micromotion \cite{devoe1989role}.

However, above a certain distance from the potential minimum that depends on the trap geometry, the anharmonicity of the potential may start to play an important role.
In general, in the absence of stochastic perturbations, the Hamiltonian phase-space for motion in rf Paul traps contains one or more approximately regular (integrable) regions. Within a regular region the motion of the ion is characterized by conserved quantities (the Hamiltonian actions, $I_j$), whose number is equal to the spatial dimension. Each trajectory is then restricted to rotations in phase-space on a manifold determined by the conserved actions, with the topology of a torus.
The time evolution is described by the angle variables $\theta_j(t) = \theta_j(0)+ \nu_j  t$ with an angular frequency $\nu_j$ that is only a function of the actions (and is time-independent), and the initial condition. Due to the micromotion, the invariant tori are periodically modulated in time within the phase-space, at the rf-drive frequency.

Beyond the integrable motion, some phase-space regions may become chaotic, and from some chaotic regions the ion can escape the trap on a very fast timescale \cite{rfchaos}. Even for chaotic motion that is bounded, the ion explores a non-zero volume in phase-space whose dimension is not reduced by conservation laws, and typically does so in an apparently random manner.
 Thus we can expect that a weak stochastic perturbation within an already chaotic  region does not change the evolution qualitatively. In the originally regular region however, a stochastic perturbation breaks the conserved quantities, changing the nature of the motion.

In this work, we consider stochastic heating and laser cooling processes in situations where the ion is initially far from the cooling limit, but within the regular phase-space parts. We strive to answer three main questions: First, due to the widespread use of a time-independent approximation of the rf trap potential (known as the pseudopotential), where can it be employed to obtain a good description of the stochastic dynamics, and where does it fail? 
Second, how do the stochastic dynamics vary with the amplitude of motion, and does the trap's anharmonicity introduce qualitative changes in comparison with harmonic motion? If it is possible to answer these two questions in a broad way, we can seek answers to the third and final question about the most advantageous values of laser cooling parameters such as detuning and intensity and also trap parameters such as electrode potentials, to efficiently load ions and to recover the ion to near the cooling limit from collisions with background gas that may put the ion in a state of high kinetic and potential energy. 

In \seq{Sec:Model} we approach the task by first transforming the Hamiltonian description of dynamics without stochastic and dissipative events to action-angle variables. 
We describe the effects of noise and cooling under the assumption that the actions change slowly due to such stochastic perturbations, as compared to the characteristic frequencies $\nu_j$. In this case, one can capture the dynamics by a Fokker-Planck equation where the actions alone suffice to describe the slow stochastic dynamics \cite{kramers1940brownian}, while the much faster rates of change of $\theta_j$ can be eliminated by averaging over these coordinates. The stochastic processes can then be characterized by action-dependent drift and diffusion rates that dynamically reshape a probability distribution over the actions as time goes on. Due to the linearity of the FP equation, different stochastic processes can be accounted for just by adding their coefficients.

In order to approximate the photon scattering during laser cooling, we employ well-established semiclassical simplifying assumptions \cite{wineland1979laser, javanainen1980,javanainen1980a,javanainen1981laser, gordon1980motion,stenholm1986semiclassical, cirac1994laser, aura2002,leibfried2003,wesenberg2007, epstein2007,marciante2010, PhysRevA.96.012519}. In \seq{Sec:Doppler} we treat photon scattering in two approximations: The coarser assumption posits that the excited state of the electron has a lifetime that is negligible on all timescales of the ion dynamics. In this case, each photon is assumed to be absorbed and then spontaneously emitted at the same point in time. We call this the ``zero lifetime'' limit and it is the limit that Javanainen and Stenholm called the ``heavy particle'' limit in their seminal work on laser cooling \cite{javanainen1980a}. The conditions of the zero lifetime limit are violated when the ion's amplitude of motion grows \cite{javanainen1981laser}, or when the driven micromotion oscillations are too fast. We remove these restrictions by treating absorption and emission as two events separated in time by intervals that are randomly picked based on the excited state lifetime. We call this the ``finite lifetime'' limit, and we restrict our treatment to a low saturation of the transition (when the ion spends most of the time in its electronic ground-state). This derivation consistently reduces to the zero lifetime limit when the lifetime of the excited state approaches zero, and our findings agree with previous results in the limits of their scope.

In the remainder of the paper, we apply this framework to study white noise heating and laser cooling with a concrete trap model. In \seq{Sec:Hamiltonian1D} we review the relevant aspects of Hamiltonian motion within Paul traps. In many trap variants \cite{leibfried2003} there is a region around the effective trap minimum where the potential is approximately a quadrupole, with a symmetry axis along which the motion corresponds to a (time-independent) harmonic oscillator, and in the transverse directions it is described by decoupled Mathieu oscillators.
Within a quadrupole potential, the heating and cooling coefficients can be approximated in closed form in some limits [\seq{Sec:Stochastic1D}]. To go beyond harmonic motion, we focus on a model surface-electrode trap with a five-wire configuration, where it is possible to separate the motion perpendicular to the electrode surface, from the other motional degrees of freedom. We restrict our attention to this 1D motion for which the potential is also strongly anharmonic in a large region, with a detailed characterization of the ranges of validity of the various approximations as a function of the action given in \seq{Sec:Numerics}.

A self-contained summary of our results and a discussion of general conclusions is presented in  \seq{Sec:Summary}, with \fig{fig:Schematic} depicting schematically the different regimes of laser cooling.
An outlook for possible applications and generalizations is laid out in \seq{Sec:Outlook}. 
To answer briefly the questions at the outset, for white noise heating we find that the pseudopotential can be safely used, which presents a significant simplification for future studies, while for laser cooling dynamics it turns out to give quantitatively and qualitatively wrong results in most scenarios far from equilibrium. 
In addition, we find that within a quadrupole potential, the efficiency of laser cooling remains independent of the amplitude of motion in the high action region, while for the strongly anharmonic potential of a surface-electrode trap we find that cooling may become heating, and also change its nature to diffusive at large amplitude motion. 
This result would be completely lacking in an approximation of the potential as a quadrupole, and even in an approximation using an anharmonic pseudopotential, serving to emphasize the importance of the interplay of nonlinearity, micromotion and stochastic dynamics in Paul traps. We also present a simple characterization of heating and cooling dynamics at different noise, laser, and trap parameters, and answer the third question posed above, for the studied model five-wire trap.
The framework that we provide allows to answer further related questions quantitatively, such as the probability or the time required for the ion to safely arrive at the cooling limit. Moreover, we find that far-from-equilibrium regions are governed by complex dynamics that hold much intrigue by themselves. 
The presented theory gives a quantitative tool for studying such dynamics and gaining better understanding of the underlying nonequilibrium mechanisms. 

\section{General model}\label{Sec:Model}

\subsection{The Hamiltonian Motion}\label{Sec:Hamiltonian}

We consider an ion of mass $m$ and charge $e$ trapped in a general Paul trap, with $\vec{r}$ and $\vec{v}\equiv \dot{\vec{r}}$ being its vector coordinate and velocity in $D=3$ dimensions, and $\vec{p}$ the canonical momentum.
 We assume that the Hamiltonian depends quadratically on the momenta, with a potential energy that is a function of the coordinates and is driven periodically in time at rf-drive frequency $\Omega$, so our starting point is the  Hamiltonian 
\be
H_0 (\vec{r},\vec{p},t)=\frac{1}{2}(\vec{p})^2+V(\vec{r},t),\quad V(\vec{r},t)=V(\vec{r},t+T),
\label{Eq:H}
\ee
with the potential having period $T=2\pi/\Omega$ (which can include the particular case where it is time-independent).
We also assume that the Hamiltonian $H_0$ can be approximated as integrable (regular) in some region of phase space, i.e.~that $D$ conserved actions exist, and the canonical action-angle coordinates $(\vec{I},\vec{\theta})$ can be defined and calculated explicitly, at least numerically \cite{rfchaos}.
A canonical transformation from the real space coordinates is then defined by
\begin{align}
&I_j=\Lambda_j(\vec{r},\vec{p},t), && \theta_j=\Theta_j (\vec{r},\vec{p},t),\label{Eq:AAtransfo}\end{align}
with the transformation functions $\Lambda_j$ and $\Theta_j$  depending explicitly on time (and being $T$-periodic), if $V(\vec{r},t)$ is.
In action-angle coordinates, the Hamiltonian $H_0$ transforms into a Hamiltonian of $\vec{I}$ alone,
and the corresponding equations of motion (with an over-dot denoting the time derivative), are
\begin{align}
&\dot{I}_j=0, && \dot{\theta}_j=\nu_j(\vec{I}),
\label{AAeom}
\end{align}
$\nu_j(\vec{I})$ being the fundamental frequencies of the motion on the torus defined by fixed actions $\vec{I}$, which are conserved and thus time-independent.

\subsection{The Fokker-Planck Equation}\label{Sec:FP}

A trapped ion is subject to different sources of noise and random perturbations \cite{brownnutt2015, turchette2000, bruzewicz2015,sedlacek2017distance}. In the absence of laser cooling, these typically include  collisions of the ion with molecules of the background gas present in the trap, fluctuations of the trap parameters (e.g. Johnson noise on voltages), and fluctuations of ambient electric fields. 
Stochastic dynamics resulting from such noise can be studied using a probability distribution in phase-space, $P(\vec{r},\vec{p},t)$, that evolves under stochastic terms in addition to the motion generated by the trap Hamiltonian, and we follow here the presentation of van Kampen \cite{kampen2007}. 
Noise heating and laser cooling with experimentally relevant parameters (discussed in more detail later) can be modelled by additive, stationary 
Gaussian white noise (approximating the noise as having no correlation after an infinitesimal time interval) with a nonzero mean. Then the evolution of $P(\vec{r},\vec{p},t)$ is described by 
  momentum drift and diffusion coefficients, $B_\alpha(\vec{r},\vec{p},t)$ and $D_{\alpha\beta}(\vec{r},\vec{p},t)$ respectively, with  $\alpha,\beta \in \{x,y,z\}$, using the FP equation 
\be
\frac{\partial P(\vec{r},\vec{p},t)}{\partial t} =\mathcal{L}_0P -\sum_{\alpha}\frac{\partial [B_\alpha P]}{\partial p_\alpha}+\frac{1}{2}\sum_{\alpha,\beta}\frac{\partial^2\left[D_{\alpha\beta} P\right]}{\partial p_\alpha \partial p_\beta},\label{Eq:FPDoppler}
\ee
where the Liouvillian $\mathcal{L}_0$ generates the Hamiltonian flow due to $H_0$ in phase-space, and is defined by
\begin{equation}
\mathcal{L}_0(\vec{r},\vec{p},t)=- \sum_\alpha \frac{p_\alpha}{m} \frac{\partial }{\partial r_\alpha}+\sum_\alpha\frac{\partial {V}\pare{\vec{r},{t}}}{\partial {r_\alpha}}\frac{\partial }{\partial p_\alpha}.
\label{Eq:Liouvillian}
\end{equation}

We now transform to the canonical action-angle coordinates $(\vec{I},\vec{\theta})$ defined in \eq{Eq:AAtransfo}.
In these variables the Liouvillian of \eq{Eq:Liouvillian} reduces to $
\mathcal{L}_0(\vec{I},\vec{\theta})=-\sum_j \nu_j(\vec{I}){\partial}/{\partial \theta_j}$.
Using the formulas of \app{Sec:generalFP}, \eq{Eq:FPDoppler} will transform to an equation in the new canonical variables. Since the Jacobian of a canonical transformation is equal to 1, the measure of $P$ is unchanged. When the timescale for change in $\vec{I}$ is much longer than the quasi-periods of rotation on the invariant torus, it is possible to approximate $P(\vec{I},\vec{\theta},t)$ by its average over the angles, ${P}(\vec{I},t)$.
 In the transformed FP equation (see in general \app{Sec:generalFP}), all terms with derivatives with respect to the angles $\vec{\theta}$  drop out when averaged over the angle. 
In the case of the rf potential, the averaging is over the entire motion at fixed action, and hence includes also an averaging over the short timescale of the rf potential. This timescale enters through the transformation functions defined in \eq{Eq:AAtransfo}. For any function $\Xi(\vec{I},\vec{\theta},t)$ of the phase space which is periodic with the rf-drive frequency,
\be \Xi\left( \vec{I},\vec{\theta},t+T\right)=\Xi\left( \vec{I},\vec{\theta},t\right),\ee 
we define the torus average denoted with an overbar,
\be
\overline{\Xi}\left( \vec{I}\right)\equiv \frac{1}{T} \int_0^{T}dt\frac{1}{(2\pi )^{ D}}\int \Xi\left(\vec{I},\vec{\theta},t\right) d^{ D}\vec{\theta}.\label{Eq:TorusAverageDef}
\ee 

 After averaging, we obtain the final form of the FP equation in action and time, that can be written in compact form using the probability flux vector $\vec{S}$ whose components are given by
\begin{align}
{S}_j(\vec{I},t)  \equiv  \Pi_j P - \frac{1}{2}\sum_k \frac{\partial }{\partial I_k} \cro{\Pi_{jk} P},
\label{Eq:FPS}
\end{align}
with the FP equation taking the form
\begin{align}
&\frac{\partial {P}(\vec{I},t) }{\partial t} = -\sum_j\frac{\partial {S_j}(\vec{I},t) }{\partial I_j} = \nonumber \\& - \sum_j\frac{\partial }{\partial I_j} \cro{\Pi_j{P}} + \frac{1}{2}\sum_{j,k}\frac{\partial^2 }{{\partial I_j}\partial I_k} \cro{\Pi_{jk}{P}} ,
\label{Eq:FPIt}
\end{align}
and the action drift and diffusion coefficients are, respectively,
\begin{align}
&\Pi_{j}(\vec{I}) =\overline{ \sum_\alpha B_{\alpha} \frac{\partial \Lambda_j}{\partial p_\alpha} + \frac{1}{2}\sum_{\alpha,\beta} D_{\alpha\beta} \frac{\partial^2 \Lambda_j}{\partial p_\alpha\partial p_\beta}}\label{Eq:FPPij},\\
&\Pi_{jk}(\vec{I}) =\overline{\sum_{\alpha,\beta} D_{\alpha\beta} \frac{\partial \Lambda_j}{\partial p_\alpha} \frac{\partial \Lambda_k}{\partial p_\beta}}.\label{Eq:FPPijk}
\end{align}
 A minimal criterion for the validity of this approximation would be a small relative change in action due to both drift and diffusion, during a cycle of the motion, i.e., 
\be \Pi_{j}(\vec{I})/ \nu_j(\vec{I}) \ll I_j, \quad \Pi_{jk}(\vec{I})/ \sqrt{\nu_j(\vec{I})\nu_k(\vec{I})}  \ll I_j I_k.\label{Eq:AdiabaticCondition}\ee 
It is clear that this adiabatic approximation breaks for $\nu_j(\vec{I})\to 0$, i.e.~close enough to a separatrix (a trajectory passing through a saddle-point in the potential energy). We return to this point later, where we calculate the stochastic coefficients in the high action regime of a surface trap. 

The FP equation can be defined for $I_j>0$, and is completely posed when boundary and initial conditions are specified. We have a reflecting boundary condition at the origin [$S(I_j=0,t)=0$], and an absorbing boundary condition ($P$ itself equals 0) could be enforced at some maximal boundary of $\vec{I}$ (corresponding to the ion escaping the trap). For an initial value problem, a typical initial condition would be
e.g.~that the ion starts at $t=0$ with some given distribution (e.g.~thermal if it is cooled or arriving from an oven, or power-law after a background-gas collision \cite{devoe2009power, rouse2017superstatistical}), or more simply, that it is approximately localized at some action value $\vec{I}$.

In the following section we consider laser cooling, that allows to counteract the effects of heating in the trap and cool the ion.

\section{Laser cooling}\label{Sec:Doppler}

In this section we derive the FP equation describing the process of laser cooling in two different semiclassical approximations. In \seq{Sec:DopplerHeavy} we review the well established derivation of the FP equation in the limit that the motion of the ion can be considered as frozen during a photon absorption-emission cycle, stating detailed conditions for the validity of this limit. In \seq{Sec:DopplerGeneral} we derive the action drift and diffusion coefficients for cooling beyond this limit, by considering the variation of the action due to photon scattering directly.

\subsection{The zero lifetime limit}\label{Sec:DopplerHeavy}

  The ion's valence electron couples the ion's center of mass motion to the electromagnetic field, while making transitions between its ground state and an excited level.
We consider the internal states of the ion in a two-level approximation and a monochromatic laser beam in a travelling wave configuration with the following  parameters: a laser wavevector ${\vec{k}}$ and wavenumber ${k}=|{\vec{k}}|$, an on-resonance Rabi-frequency ${\Omega}_{\rm{R}}$ (with its squared magnitude proportional to the laser intensity), and a laser frequency ${\omega}_{\rm{L}}$ detuned by ${ \Delta}$ from the resonant electronic transition, whose linewidth (inverse lifetime) is ${\Gamma}$.
 For each absorption or emission process, the atom suffers a recoil of magnitude ${p}_{\rm{r}}={\hbar} {k}$ which is parallel (anti-parallel) to the momentum of the absorbed (emitted) photon, and ensures conservation of the total momentum (${\hbar}$ being Planck's constant). 

For optical transitions and nonrelativistic ion velocities ${\vec{v}}$ (much smaller than the speed of light ${c}$), the same parameter ${k}$ can be used for photons resonant with the internal electronic transition, for the detuned laser photons, and for Doppler-shifted photons, i.e.,
\be {k}=\frac{{\omega}_{\rm{L}}}{{c}}\approx \frac{{\omega}_{\rm{L}}+{\Delta}}{{c}}\approx \frac{{\omega}_{\rm{L}}-{\vec{k}}\cdot {\vec{v}}}{{c}}.\label{Eq:kapprox}\ee
 A stochastic laser cooling process can be described in a semiclassical approximation by using the FP equation for the probability distribution $P(\vec{r},\vec{p},t)$ of the ion's center of mass coordinates alone. 
 When the requirements of the derivation are fulfilled  (with the exact conditions discussed below), which can be qualified schematically as the limit of slow enough motion at low amplitude within the pseudopotential approximation, the ion motion can be assumed as frozen during the lifetime of the excited level (in \cite{javanainen1980a} this was called the heavy particle limit). The FP equation in this limit, takes the form of \eq{Eq:FPDoppler}, with two functions that we denote by $B_\alpha^{\rm z}(\vec{p})$ and $D_{\alpha\beta}^{\rm z}(\vec{p}) $.

 The vector function $B_\alpha^{\rm z}(\vec{p})$ gives the mean momentum transfer rate due to the radiation pressure force, that acts on the ion as a function of its momentum. The mean momentum gain per cycle is just $p_r$ of the absorbed photon, since the probability to emit a photon in a certain direction is invariant under inversion of the coordinates. 
The rate of absorption-emission cycles  is given by $\Gamma$ times the occupation probability of the electron in the excited level, for an ion having momentum $\vec{p}$ in phase-space:
\be\rho_s(\vec{p}) = \frac{s/2}{1+s+\pare{2\Delta_{\rm{eff}}/\Gamma}^2}.\label{Eq:rho_p_z}\ee
Here the saturation parameter $s$ is defined by
\be s=2\left(\abs{{\Omega}_{\rm{R}}}/{\Gamma}\right)^2,\ee
and using the relation of the phase-space momentum to the velocity,
 \be \vec{v}=\vec{p}/m,\label{Eq:vp}\ee
 we have that due to the Doppler shift, the detuning of the laser relative to the ion resonance at rest, in the frame of reference of the ion, is
\be \Delta_{\rm{eff}}=\Delta-\vec{k}\cdot\vec{v}.\ee
 Thus $\rho_s(\vec{p})$ describes a Lorentzian with velocity center and velocity half-width given respectively (along the direction of $\vec{k}$)  by
\be v_0=\Delta/k,\qquad\delta v=\Gamma/k.
\label{Eq:Lorentzian}\ee
We note that $\rho_s(\vec{p})<1/2$, which expresses the saturation of the two-level system for large values of $s$. With $\hat{k}_\alpha=\vec{k}_\alpha/k$ to specify the direction of the laser wavevector,  the momentum change rate is given by
\be B_\alpha^{\rm z}(\vec{p})=p_{\rm{r}} \hat{k}_\alpha\Gamma\rho_s(\vec{p}).\label{Eq:BHeavy}\ee

The tensor function $D_{\alpha\beta}^{\rm z}(\vec{p})$ describes diffusion in momentum  due to two sources -- the variance in photon absorption, and the spontaneous emission. The second moment of the angular distribution of emitted photons, assumed to be a dipolar transition \cite{javanainen1980}, is for a linearly polarized beam,
\be \mu_{\alpha\beta}=\frac{2}{5}\delta_{\alpha\beta}-\frac{1}{5}\hat{e}_\alpha\hat{e}_\beta,\label{Eq:mu}\ee
(with $\hat{e}$ the polarization unit vector). {The assumed linear polarization can be replaced by circular polarization with just the values of $\mu_{\alpha\beta}$ changing appropriately. This tensor enters as a prefactor in the mean diffusion per cycle due to spontaneous emission, with the cycles proceeding at the rate $\Gamma\rho_s(\vec{p})$.} 
In addition, since the photon absorption events are discrete, there is an independent contribution proportional to the variance of their number per unit time \cite{stenholm1986semiclassical}. Assuming uncorrelated Poissonian photon statistics, the variance in the number of absorptions is equal to their mean number, multiplied by a term proportional to $\hat{k}_\alpha\hat{k}_\beta$ accounting for the well-defined direction of momentum transfer, for a total momentum diffusion coefficient
\be 
D_{\alpha\beta}^{\rm z}(\vec{p})=p_{\rm{r}}^2 \left(\hat{k}_\alpha \hat{k}_\beta+\mu_{\alpha\beta}\right) \Gamma \rho_s(\vec{p})\label{Eq:DHeavy}.\ee

Plugging $B_\alpha^{\rm z}$ of \eq{Eq:BHeavy} and $D_{\alpha\beta}^{\rm z}$ of \eq{Eq:DHeavy} into \eqs{Eq:FPPij}-\eqref{Eq:FPPijk} gives the coefficients for the angle-averaged FP equation [\eq{Eq:FPIt}] of cooling in the zero lifetime limit,
 \begin{align} &\Pi_{j}^{\rm z}(\vec{I}) =\overline{ \sum_\alpha B_{\alpha}^{\rm z} \frac{\partial \Lambda_j}{\partial p_\alpha} + \frac{1}{2}\sum_{\alpha,\beta} D^{\rm z}_{\alpha\beta} \frac{\partial^2 \Lambda_j}{\partial p_\alpha\partial p_\beta}}\label{Eq:Pijh} \\ &\Pi^{\rm z}_{jk}(\vec{I}) =\overline{\sum_{\alpha,\beta} D^{\rm z}_{\alpha\beta} \frac{\partial \Lambda_j}{\partial p_\alpha} \frac{\partial \Lambda_k}{\partial p_\beta}}.\label{Eq:Pijkh}\end{align}

The derivation of the FP equation \cite{javanainen1980a} assumes an expansion in the following small parameter,
\be  k{{{p}_{\rm{r}}}}/{(m\Gamma)} \ll 1 \iff 
E_{\rm recoil}={p}_{\rm{r}}^2/(2 m) \ll { \hbar}{ \Gamma},\label{Eq:FPHeavyParameter}\ee
implying that the relative change of $P(\vec{r},\vec{p},t)$ in each cycle is small, due to the smallness of the recoil momentum and the associated kinetic energy ($E_{\rm recoil}$), with respect to the scale determined by the Lorentzian due to the width of the excited level. In addition, the ion's internal (electronic) degrees of freedom have  to be adiabatically eliminated. The rate of decay of the excited state (${\Gamma}$) must be faster than the unperturbed evolution of $P(\vec{r},\vec{p},t)$ due to the Liouvillian of \eq{Eq:Liouvillian}, i.e. we require
\be \left|p_\alpha\frac{\partial P}{\partial r_\alpha}\right| \ll \Gamma P, \qquad \left|\frac{\partial V}{\partial r_\alpha}\frac{\partial P}{\partial p_\alpha}\right| \ll \Gamma P.\label{Eq:Padiabatic}\ee
  For motion within a 3D harmonic potential,
assuming a single length scale of variation for $P$ with $r_\alpha$, determined by the amplitude of the motion in the harmonic limit, the first condition in \eq{Eq:Padiabatic} amounts to 
\be \nu_j\ll\Gamma \label{Eq:Unresolved},
\ee
known as  the unresolved sideband limit.
The second condition in \eq{Eq:Padiabatic} can be written as 
\be {\rm max}\{\dot{\vec{v}}\cdot \vec{k}\} \ll \Gamma^2,\label{Eq:Fast}\ee
by using the fact that the largest variation of $P$ with momentum results from absorption at the velocity range $\delta v$ (around $v_0$), defined in \eq{Eq:Lorentzian}. Thus the condition in \eq{Eq:Fast} limits the validity of the treatment to low amplitude and low velocity motion in a harmonic oscillator. 

For motion in the time dependent rf potential, the averaging over the torus takes care of the rf modulated {velocity} along the trajectory due to the micromotion, which modifies the instantaneous Doppler shift. Equation \eqref{Eq:Padiabatic} still gives  conditions for the validity of the {current} treatment, which must be amended with 
\be  \Omega/2\pi\ll {\Gamma}.\label{Eq:OmegallGamma}\ee
The conditions that limit the treatment to a small amplitude of motion, and the limitation to the unresolved sideband limit and a small rf frequency in \eq{Eq:OmegallGamma} are very restricting, in particular with  state-of-the-art-traps.
 In the next subsection we develop the semiclassical laser cooling theory in the limit of finite lifetime  of the excited electronic state.

\subsection{Laser cooling in the finite lifetime limit}\label{Sec:DopplerGeneral}

When either of {the conditions} \eq{Eq:Unresolved}, \eq{Eq:Fast} or \eq{Eq:OmegallGamma} do not hold, the ion motion can no longer be considered {to be} frozen during the absorption-emission cycle, and within the lifetime of the excited level ($1/\Gamma$), the ion's velocity and position may change significantly.
A Fokker-Planck equation in energy space has been derived for the large amplitude motion of an ion in a 1D harmonic oscillator potential \cite{javanainen1981laser}, by making explicit use of  harmonic oscillator wavefunctions. We  now  develop a semiclassical derivation \cite{wineland1979laser,wesenberg2007,PhysRevA.96.012519}, that allows to generalize the results of the previous subsection to the motion in any trap potential, including micromotion, at any amplitude.

The basic principle of this semiclassical derivation is to assume conservation of energy and momentum at each absorption and at each emission event separately, accounting for the ion's centre of mass and internal coordinates together with the photon.
Between the absorption and emission, we assume that the ion moves completely classically and is decoupled from the electromagnetic field. 
It is worth noting that this treatment only alters the description of the emission events, while absorption events are effectively equivalent to their zero-lifetime description. 
As we will see below,
the validity of this treatment requires a low laser intensity, i.e.
\be s\ll 1.\label{Eq:sll1}\ee
With that condition, the  probability of an ion to absorb a photon at a given point in phase-space in a small time interval $dt$ equals $\Gamma\rho dt$, with $\rho$ obtained from $\rho_s$ of \eq{Eq:rho_p_z} by expansion in $s$,
\be\rho(\vec{p}) = \frac{s/2}{1+\pare{2\Delta_{\rm{eff}}/\Gamma}^2}, \qquad \Delta_{\rm eff}=\Delta-\vec{k}\cdot \vec{v},\label{Eq:rho_p_z2}\ee
and $\vec{v} = \vec{p}/m$ [\eq{Eq:vp}]. The resulting absorption and emission rates are equal on average, even if there are finite delays between these events.
A more detailed derivation of the photon absorption probability from the optical Bloch equations will be presented separately in \cite{rfdist} using a Floquet approach, which allows one to improve the accuracy of accounting for the micromotion drive.

Let us consider an ion moving in the trap, and at some arbitrary time $t_a$ when it is at position $\vec{r}_a$ with momentum $\vec{p}_a$ (velocity $\vec{v}_a$), it absorbs a laser photon of energy  $\hbar(\omega_{\rm L}+\Delta)$. 
Due to the absorption, the ion's momentum changes by $p_r{\hat{k}}$, and the electron is excited by the Doppler shifted photon, consistent with the level width $\Gamma$ and the Lorentzian of \eq{Eq:rho_p_z}, and the condition in \eq{Eq:kapprox}. To simplify the notation below, we define the phase-space point
\be Z_a\equiv\{\vec{r}_a,\vec{p}_a,t_a\}.\label{Eq:Za}\ee
The change in each action $I_j$ due to the absorption is then, to second order in the recoil momentum,
\bem \delta I_{j}^{(a)}=I_{j} (\vec{r}_a,\vec{p}_a +p_{\rm r}{\hat{k}},t_a) -I_{j}(\vec{r}_a,\vec{p}_a,t_a) \approx \\  p_{\rm r}\sum_\alpha {\hat{k}}_\alpha\frac{\partial \Lambda_j(Z_a)}{\partial p_\alpha} +\frac{1}{2} p_{\rm r}^2\sum_{\alpha,\beta} {\hat{k}}_\alpha {\hat{k}}_\beta \frac{\partial^2 \Lambda_j(Z_a)}{\partial p_\alpha\partial p_\beta}.\label{Eq:deltaIa}\end{multline}
In \eq{Eq:deltaIa}, $\delta I_{j}^{(a)}$ is a random variable, that is conditioned on the absorption taking place at $Z_a$.

Continuing, the ion moves under the influence of the trap and at time $t_e$ reaches the phase space point $\{\vec{r}_e,\vec{p}_e\}$. We take this point to be 
just the position and momentum that the ion would have reached, if it hadn't absorbed a photon, due to the Hamiltonian evolution from $t_a$ to $t_e$ on the torus of fixed $\vec{I}$. Hence we neglect the small change for the initial condition at $t_a$ due the ion having made a step $p_{\rm r}{\hat{k}}$, which is just the approximation at the basis of the FP approach, that a lot of stochastic events are required to cause a significant change in the action distribution. 
Now at time $t_e$ the ion spontaneously emits a photon with momentum $\hbar\vec{\kappa}\approx p_{\rm r}{\hat{\kappa}}$, where the condition of \eq{Eq:kapprox} has been used. 
 The change in action due to emission occurring at $Z_e$, of a photon propagating along $\hat{\kappa}$, and conditioned on absorption at $Z_a$,  is
\bem \delta I_{j}^{(e)}=I_{j}(\vec{r}_e,\vec{p}_e-p_{\rm r}\hat{\kappa},t_e)- I_{j}(\vec{r}_e,\vec{p}_e,t_e) \approx\\  -p_{\rm r}\sum_\alpha \hat{\kappa}_{\alpha}\frac{\partial \Lambda_j(Z_e)}{\partial p_\alpha} +\frac{1}{2} p_{\rm r}^2\sum_{\alpha,\beta} \hat{\kappa}_{\alpha}\hat{\kappa}_{\beta} \frac{\partial^2 \Lambda_j(Z_e)}{\partial p_\alpha\partial p_\beta}.\label{Eq:deltaIe}\end{multline}

Summing \eqss{Eq:deltaIa}{Eq:deltaIe}, the total change to the action given that the absorption occurred at $Z_a$ and the emission at $Z_e$ with the photon going along $\hat\kappa$, is
\be \delta  I_{j}  = \delta I_{j}^{\left(a\right)} + \delta I_{j}^{\left(e\right)}.\label{Eq:deltaI0}\ee
To recap, the ion drift and diffusion in action is due to three sources of randomness: the time and phase-space point of the absorption event, the time and phase-space point of the emission event  through decay from the excited level with finite lifetime of $1/\Gamma$, and the direction  $\hat{\kappa}$ of the emitted photon. 
We assume that when coarse-grained over many absorption-emission cycles, the action  evolves in small deviations $\delta \vec{I}$ [as in \eq{Eq:deltaI0}] about the given torus $\vec{I}$, due to the accumulation of many small action kicks. The mean action drift and diffusion rates have to be calculated by averaging over multiple emission-absorption cycles using the distribution of the action increments. To a good approximation, the distribution of the sum of many such small action kicks may be taken to be Gaussian due to the central limit theorem. Here we can invoke \eq{Eq:sll1} and neglect any 
correlation between the absorption and emission coming from the fact that absorption
is impossible while the ion is in the excited level, which can only contribute at order $s^2$, and similarly for stimulated emission processes that can be neglected to lowest order in $s$.
In the following we calculate the first and second moments of the distribution of action kicks, on the torus $\vec I$ of the motion.

 We start with the random variables of the emission.
The mean change in action due to emission is obtained by taking the expectation value of \eq{Eq:deltaIe} over the two random variables of emission; the direction of the emitted photons and the phase-space point they are emitted at.
The mean over the spherical distribution of $\hat{\kappa}$ is denoted 
by $\langle \cdot\rangle_{\hat{\kappa}}$, and we have 
\be \langle \kappa_\alpha \rangle_{\hat{\kappa}} = 0, \qquad \langle
{\kappa}_{\alpha}
{\kappa}_{\beta}\rangle_{\hat{\kappa}}=\mu_{\alpha\beta},\ee 
 in which the term linear in $\hat{\kappa}$ components vanishes since it is reflection-invariant, and the second moment is defined by the tensor $\mu_{\alpha\beta}$ in \eq{Eq:mu}. The emission time is also a random variable with an exponential distribution for decay from the excited level. Given an absorption that occurred at $Z_a$, the mean value of any function of phase-space at the time of emission, averaged over the random times of emission, denoted by $\langle \cdot\rangle_\Gamma$, is 
\be\left \langle \Xi(Z_a) \right\rangle_{\Gamma} \equiv  \int_0^\infty
\Gamma e^{-\Gamma t'} \Xi(Z(t_a+t';
Z(t_a)=Z_a))dt'.\label{Eq:GammaDecay}\ee 
The time integral is to be performed along the trajectory $Z(t_a+t')$ in the notation of \eq{Eq:Za}. 
Therefore we get for the emission
\be \langle \delta I_{j}^{(e)}\rangle_{\hat{\kappa},\Gamma} =
\frac{1}{2} p_{\rm r}^2\sum_{\alpha,\beta} \mu_{\alpha\beta}
\left\langle\frac{\partial^2 \Lambda_j(Z_a)}{\partial p_\alpha\partial
    p_\beta} \right\rangle_\Gamma.\label{Eq:deltaIe2}\ee 

We consider now the effect of the randomness of the absorption event, conditioned so far to occur at the phase-space point and time given by $Z_a$. 
As discussed above, the  probability of the ion to absorb a photon at a given point in phase-space in a time interval $dt$ equals $ \Gamma\rho dt$ [\eq{Eq:rho_p_z2}]. The mean of any random process that depends on absorption at $Z_a$, calculated for motion during a time interval $\delta t$, can be obtained from
\be {\left\langle {\Xi}\right \rangle}_{\delta t}\equiv  \int_0^{\delta t} \Xi\left(Z_a(t_a)\right)\Gamma\rho\left(Z_a(t_a)\right)dt_a,\ee
where $\delta t $ is the intermediate timescale over which many absorption-emission cycles occur during a large number of rotations on the torus, while the action variation remains small.
We assume ergodicity for motion on the torus -- i.e.~that the average over many photon scattering events during many rotations on the torus, is equal to the average over the torus angles (and the relative micromotion phase), defined in \eq{Eq:TorusAverageDef}. This ergodicity assumption, over a timescale given by $\delta t$ is given by writing for the rate of any random process
\be \frac{1}{\delta t} \left\langle {\Xi}\right \rangle_{\delta t} = \Gamma \overline{\rho\Xi},\label{Eq:Ergodic}\ee
with the torus average operation 
\be
\Gamma \overline{\rho\Xi}\equiv \frac{1}{\pi} \int_0^{\pi}dt\frac{1}{(2\pi )^{ D}}\int \Gamma\rho\left(Z_a\right) \Xi\left(Z_a\right) d^{ D}\vec{\theta}.\label{Eq:TorusAverageDef2}
\ee 

Averaging over the absorption events allows us to obtain the FP equation [\eq{Eq:FPIt}] for $P(\vec{I},t)$, evolving adiabatically at a timescale longer than that for multiple emission-absorption cycles. The first two moments of the action deviations determine the FP equation coefficients \footnote{see Chapter VIII of \cite{kampen2007}.},
\be \Pi_j^{\rm f}(\vec{I})=\langle\delta I_j\rangle/\delta t,\qquad \Pi_{jk}^{\rm f}(\vec{I})=\langle\delta I_j \delta I_k\rangle/\delta t,\ee
where $\langle\cdot\rangle$ denotes the ensemble average to first order in $\delta t$, i.e.~$\langle\cdot\rangle=\langle\cdot\rangle_{\hat{\kappa},\Gamma,\delta t}$. 
Hence we get the FP coefficients
\be
\Pi_{j}^{\rm f}(\vec{I}) =\Gamma\overline{ \rho(Z_a)\langle\delta I_j\rangle_{\hat\kappa,\Gamma}} \label{Eq:FPPci},\ee
and
\be  
\Pi_{jk}^{\rm f}(\vec{I}) =\Gamma\overline{ \rho(Z_a)\langle  \delta I_j \delta I_k\rangle_{\hat\kappa,\Gamma}},\label{Eq:FPPcij}
\ee
where we find using \eqss{Eq:deltaI0}{Eq:deltaIe2}, 
\bem \langle\delta  I_{j} \rangle_{\hat{\kappa},\Gamma}  =
p_{\rm r} \sum_\alpha {\hat{k}}_\alpha\frac{\partial
  \Lambda_j(Z_a)}{\partial p_\alpha} \\+ \frac{1}{2} p_{\rm r}^2
\sum_{\alpha,\beta}\left[  {\hat{k}}_\alpha {\hat{k}}_\beta
  \frac{\partial^2 \Lambda_j(Z_a)}{\partial p_\alpha\partial
    p_\beta}+\mu_{\alpha\beta} \left\langle\frac{\partial^2
      \Lambda_j(Z_a)}{\partial p_\alpha\partial p_\beta}
  \right\rangle_\Gamma  \right],\label{Eq:deltaI1}\end{multline} 
and to order $p_r^2$ and $s$,
\bem \langle \delta I_j \delta I_k \rangle_{\hat\kappa,\Gamma} =\\ p_r^2
      \sum_{\alpha,\beta}\left[\hat{k}_\alpha\hat{k}_\beta
        \frac{\partial\Lambda_j}{ \partial p_\alpha}
        \frac{\partial\Lambda_k}{\partial p_\beta} + \mu_{\alpha\beta}
        \left\langle\frac{\partial\Lambda_j}{ \partial p_\alpha}
          \frac{\partial\Lambda_k}{\partial p_\beta}
        \right\rangle_\Gamma \right] ,\label{Eq:Sigma2}\end{multline} 
        in which, as in \eq{Eq:deltaI1},
all terms are
      functions of $Z_a$, the phase-space time and point where the
      absorption occurred, on a given torus with actions $\vec{I}$.

  The expressions in \eqs{Eq:FPPci} and \eqref{Eq:FPPcij} hold for an arbitrary rf potential, with the conditions in  \eq{Eq:FPHeavyParameter} and \eqref{Eq:sll1} assumed in the derivation, in addition to the adiabaticity conditions of \eq{Eq:AdiabaticCondition}. Assuming that the timescale separation (expressed by these adiabaticity conditions) is justified, the only foreseeable case where the ergodicity assumption of \eq{Eq:Ergodic} may fail to hold is the case of one or two decoupled degrees of freedom of the motion which are directed transversally to the laser $\vec k$-vector. Close to such a degenerate case, e.g.~when there are two nearly degenerate modes of oscillation, the dynamics have to be treated in more detail (such as in \cite{javanainen1980}, for the case of  time-independent harmonic oscillators). The absorption probability taken in \eq{Eq:rho_p_z2} can be improved as in \cite{rfdist} or generalized to a more complicated setup or level-structure. 

  The difference of the derived action drift and diffusion coefficients to their expressions within the zero lifetime limit [in \eqss{Eq:Pijh}{Eq:Pijkh}], lies  in the integration over the waiting time distribution for spontaneous emission from the excited electronic state. As a basic consistency check, we see that whenever we can assume an instantaneous absorption-emission cycle and write $\Gamma e^{-\Gamma t'}\approx \delta(t')$ in $\langle \cdot\rangle_\Gamma $ of \eq{Eq:GammaDecay}, the drift and diffusion coefficients reduce in form to those calculated for the zero lifetime limit (and under the assumption $s\ll 1$), whence $\Pi_j^{\rm f}\to \Pi_j^{\rm z}$ and $\Pi_{jk}^{\rm f}\to \Pi_{jk}^{\rm z}$.
The cooling coefficients can be further simplified in some limits, and in some cases even integrated in closed form, as we show in the following.

\section{Hamiltonian motion in 1D}\label{Sec:Hamiltonian1D}

\subsection{Anharmonic trap potential}\label{Sec:rfpotential}

\begin{figure}
\includegraphics[width=3.3in]{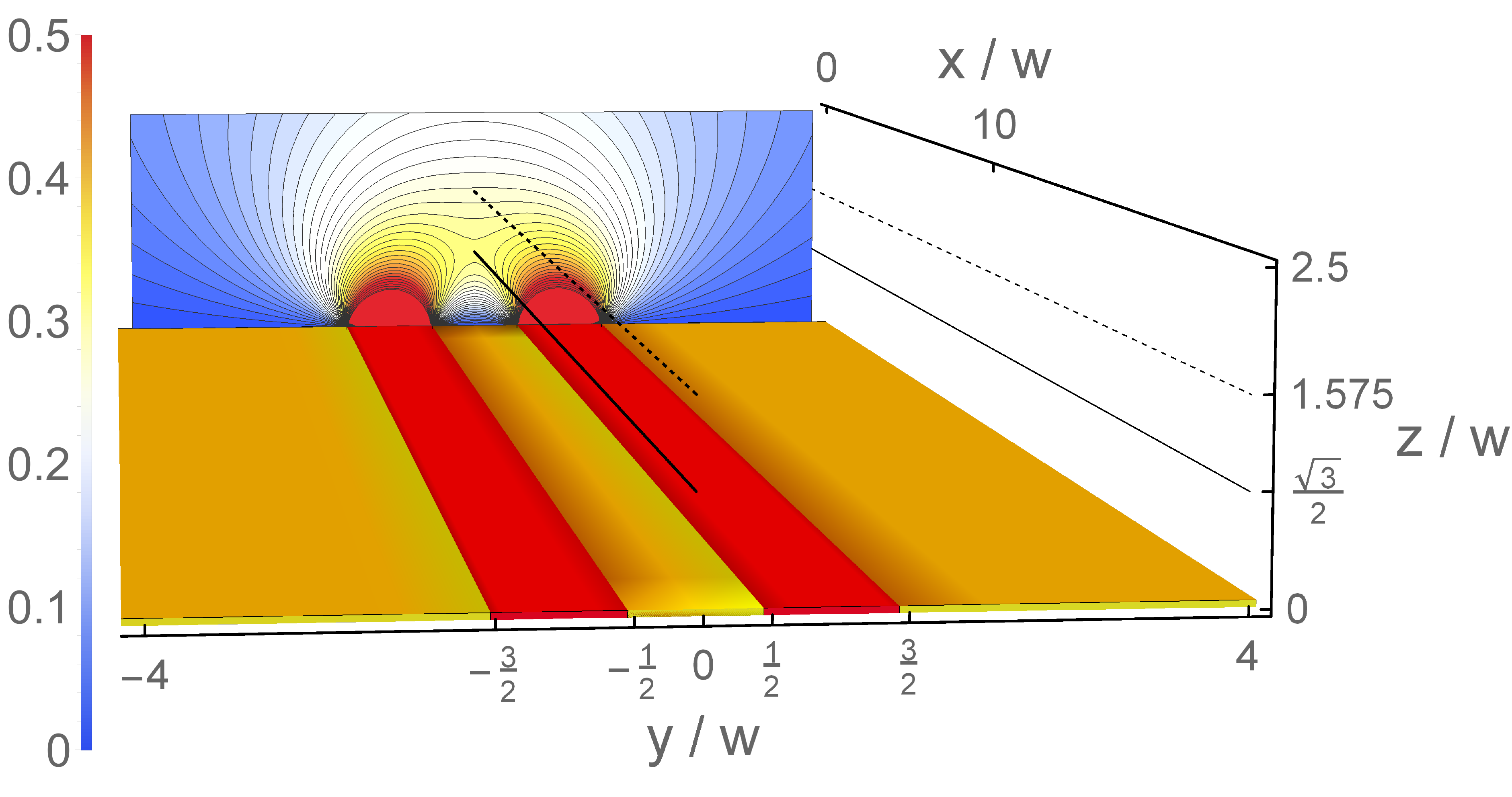}
\caption{Layout and rf-potential of the five-wire surface trap. All electrodes lie in the $z=0$ plane, with the two electrodes connected to the rf-drive shown in red. They are of width $w$ in the $y$-direction with their center offset by $\pm w$ from $y=0$. Along the $x$-direction they are approximated as having an infinite extent. The remainder of the $z=0$ plane is filled by grounded surfaces, shown in  gold. Setting the rf-electrodes to $1\,$V produces the equipotential lines in the plane $x=0$ shown as a contour plot in the back of the figure, with the bar legend showing the potential in units of V. The thick solid black line shows the potential minimum line ($z=z_s$) and the thick dashed line shows the saddle-line of the potential.}
\label{fig:trap}
\end{figure}

For a general anharmonic and time-dependent rf potential $V(\vec{r},t)$, the motion has to be solved numerically. In \cite{rfchaos} the Hamiltonian motion of an ion in the 3D potential of a model surface-electrode trap has been treated for a broad range of parameters, and a range of parameters has been identified wherein the Hamiltonian motion within the full time dependent rf potential is well described by the (time-independent) pseudopotential approximation, with a phase space  which is to a high degree regular. 
To clearly illustrate the main results that we derive here, we present numerical calculations for motion in one spatial dimension, orthogonal to the electrode surface, along the $z$ axis (see \fig{fig:trap}). In the five-wire trap in a symmetric configuration (without DC coupling of the radial $yz$ motion), the $z$ motion (which is both time-dependent and nonlinear) decouples exactly for the initial conditions $y=\dot{y}=0$, for which $y(t)=0$ at all times.
The canonical momentum is $p_z=mv_z$, with the velocity $v_z=\dot{z}$.

From this point on we use nondimensional units, obtained by rescaling the time $t$ by half the micromotion frequency, (we choose $\Omega=2\pi\times 100\,{\rm MHz}$), and measuring distances using a natural length scale of the problem, the width of each of the two rf electrodes (we take $w=50\,{\rm \mu m}$);
\begin{equation}
 z \to z/w, \qquad t \to \Omega t /2.
\label{Eq:rescaling}
\end{equation}
 The details of this rescaling (which makes all physical quantities and parameters nondimensional), including the values of the parameters chosen for the numerical calculations, are summarized in \app{Sec:Nondim}. With this rescaling, the rf-drive frequency becomes $\Omega=2$ and its period $T=\pi$. We also set the ion's mass and charge $m=1$ and $e=1$, absorbing their values in the parameters of the nondimensional 1D rf potential, given by
\be V_{\rm{rf}}^{\rm 1D}(z,t) = V_0(z)+ V_2(z)\cos 2t,\label{1Ddimrfpot}\ee
with 
\bea &V_0(z)=\frac{1}{2} a_z{\pare{z-z_s}}^2,\label{Eq:V0} \\
 &V_2(z)=-\frac{4}{\pi}q_5 \cro{\arctan(\frac{1}{2z})-\arctan(\frac{3}{2z})}.
\label{Eq:V2} \end{align}
The nondimensional parameters $a_z$ and $q_5$ are determined by the electrode voltages and geometry (and the ion's charge to mass ratio), {see \eq{aq5def}}. The potential $V_2$ of \eq{Eq:V2} vanishes at the saddle-point $z=\sqrt{3}/2$, and hence setting $z_s=\sqrt{3}/2\approx 0.866$ in $V_0$ of \eq{Eq:V0}, makes $V_{\rm{rf}}^{\rm 1D}(z,t) $ vanish for any value of $t$ at $z_s$, which then forms the center of the trapping region (see below).
The action-angle coordinates $(I,\theta)$, in the integrable approximation of the motion, can be calculated using a stroboscopic map of the motion taken at times  $t \pmod{\pi}$, where the action is then related to the phase-space area $J$ bounded within an invariant curve (\cite{rfchaos}), by \be I=J/2\pi.\label{Eq:Action}\ee 

\begin{figure}[!t]
\includegraphics[width=3.3in]{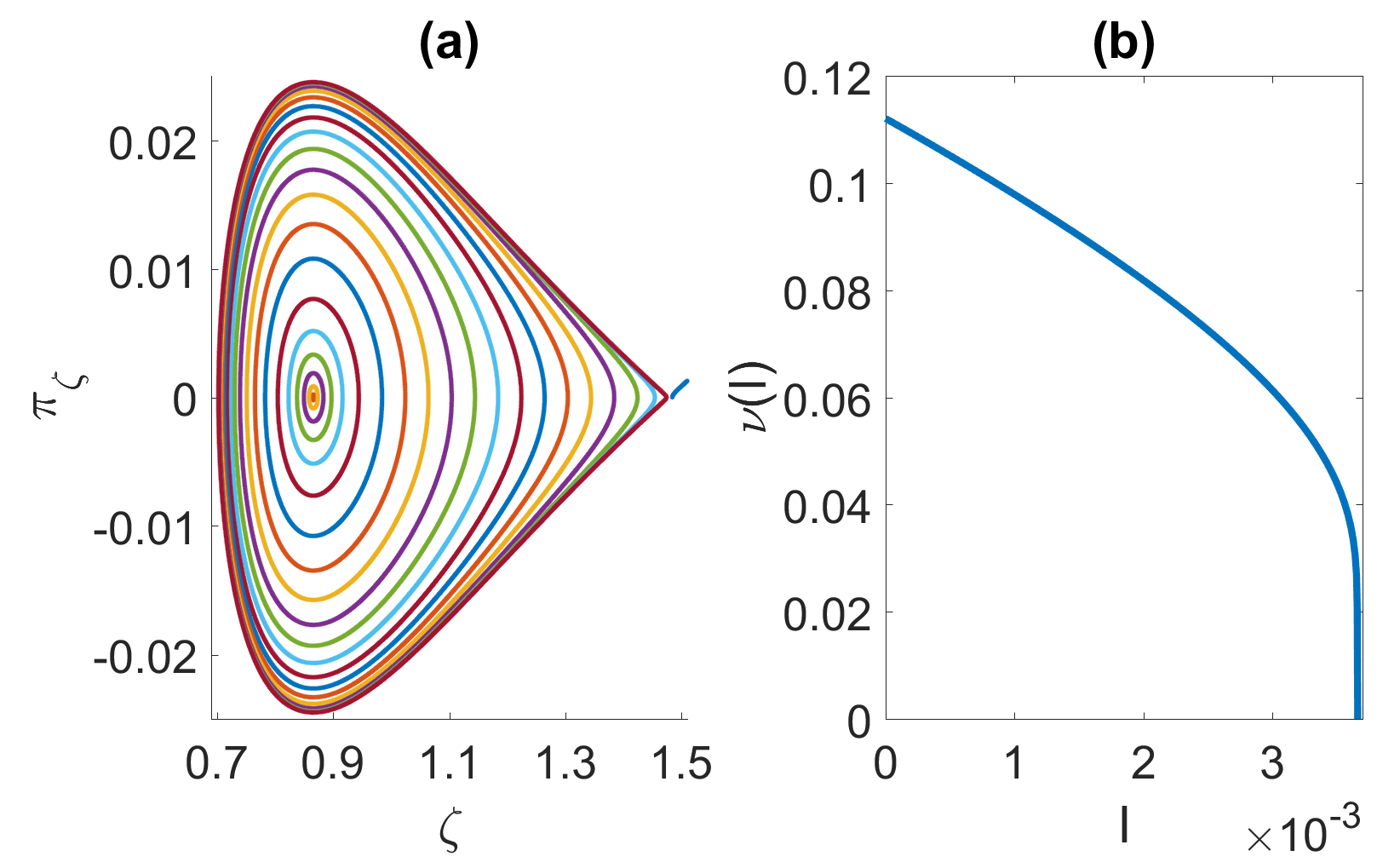}
\caption{(a) The phase space of 1D motion perpendicular to the electrodes plane of a five-wire trap, in the pseudopotential approximation [\eq{2paramSurfacePseudoPot1D}], with $\zeta$ and $\pi_\zeta$ being the coordinate and conjugate momentum in nondimensional units (see \eq{Eq:rescaling} and \app{Sec:Nondim}). The stable fixed point $\zeta_s\approx 0.866$ forms the center of the island, and the  unstable fixed point $\zeta_u\approx 1.48$ sits at the ``tip'' of the bounded part of phase space, with the curve passing through it forming the separatrix. (b) The frequency $\nu(I)$ of rotation around the invariant curves of the phase space, as a function of the action $I$. We see that $\nu(I=0)= \nu_z\approx 0.112$, and a close examination shows that only very close to the maximal action, which corresponds to the separatrix going through the unstable fixed point, the frequency sharply goes to 0, i.e.~$\nu(I)\to 0$  (this happens within a segment of width $\delta I\lesssim 0.01\times 10^{-3}$, about $I_{\rm max}\approx 3.67\times 10^{-3}$).
}\label{fig:ps}
\end{figure}

The pseudopotential approximation to \eq{1Ddimrfpot}  can be obtained by a time
dependent canonical transformation from the 
phase space variables $\{z,p_z,t\}$,
to new canonical variables that we denote as  $\zeta$ for the coordinate and $\pi_\zeta$ for the momentum. The transformation is a perturbative expansion to second order in $\nu$, the ratio of the secular frequency to the rf frequency, assuming that $V_0\sim\nu^2$ and $V_2\sim\nu$, and results in a time-independent Hamiltonian, with the time-averaged pseudopotential given by \cite{rfchaos} 
\begin{equation}
{V}_{\rm{ps}}^{1{\rm D}}(\zeta)=a_z \frac{(\zeta-\zeta_s)^2}{2}+q_{5}^2\frac{16 \left(3-4\zeta^2\right)^2}{\pi^2\left(9+40\zeta^2+16\zeta^4\right)^2}.
\label{2paramSurfacePseudoPot1D}
\end{equation}
The pseudopotential has two fixed points, where the field vanishes.
 The first point, $\zeta_s=z_s={\sqrt{3}}/{2}\approx 0.866$, is the stable fixed point at the center of the trap, whose value is independent of the parameters $a_z$ and $q_5$, within the pseudopotential (as well as the rf potential). The second point, $\zeta_u$, is an unstable fixed point, a local maximum of the pseudopotential, beyond which the ion escapes the trap (see \fig{fig:ps}).
The canonical transformation relating the pseudopotential coordinates $\{\zeta,\pi_\zeta\}$ to the original coordinates $\{z,p_z\}$ at every point of phase-space is, to the leading order in the expansion,
\be \zeta = z,\qquad \pi_\zeta = p_z + \frac{1}{2}\sin(2t)V_2'(z). \label{Eq:TransPseudo}\ee

For our numerical study, although obtaining the transformation functions of \eq{Eq:AAtransfo} to the action-angle coordinates within the 1D rf potential is numerically accessible (as demonstrated in \cite{rfchaos}), obtaining smooth partial derivatives [required in \eqs{Eq:FPPij}-\eqref{Eq:FPPijk}] is numerically more demanding. Hence we  approximate the 1D transformation $I=\Lambda(z,{p_z},t)$ using the pseudopotential approximation. 
Introducing explicit subscripts to denote the form of the potential within which $\Lambda$ is calculated, we derive in \app{App:CanonicalDerivatives} the approximate relations
\bea \frac{\partial \Lambda_{\rm rf}(z,p_z,t)}{\partial p_{z}}&\approx \frac{\partial \Lambda_{\rm ps}(\zeta,\pi_\zeta)}{\partial \pi_\zeta} =\frac{\pi_\zeta}{\nu(I)},\nonumber\\
 \frac{\partial^2 \Lambda_{\rm rf}(z,p_z,t)}{\partial p_{z}^2}& \approx \frac{\partial^2 \Lambda_{\rm ps}(\zeta,\pi_\zeta)}{\partial \pi_\zeta^2}=\frac{1}{\nu(I)}-\frac{\pi_\zeta^2}{\nu(I)^3}\frac{d\nu(I)}{dI}.\label{Eq:TransPseudoPartial}\end{align}
The corrections to \eq{Eq:TransPseudoPartial} for the case of the rf potential will be of a higher (at least second) order in $\nu$, and hence are expected to be quantitatively small. 

\subsection{Quadrupole trap potential}\label{Sec:Linearization}

 As discussed in the introduction, in many cases the Paul trap potential can be approximated as a quadrupole potential around an effective minimum of the trap. Often the motion along a certain direction is nearly decoupled from the other directions due to the symmetry of the trap electrodes. Depending on the nature of the potential in this direction, it can be described by a time-dependent Mathieu oscillator if the potential is periodically modulated in time, or a harmonic oscillator if the potential is static (time-independent). For concreteness, we consider the idealized five-wire surface electrode trap introduced in \seq{Sec:rfpotential} in the following. This geometry naturally implements the harmonic oscillator along the trap axis of symmetry ($x$, not considered here), and Mathieu oscillators in the transverse plane. For simplicity we here consider motion along one direction (denoted by $z$) and compare a Mathieu oscillator and a harmonic oscillator with identical secular frequencies.
  
The leading order expansion of ${V}_{\rm{rf}}^{1{\rm D}}$ of \eq{1Ddimrfpot} about $z_s$ gives the Mathieu oscillator potential
\be {V}_{\rm{rf}}^{1{\rm D}} \to {V}_{\rm{M.o.}}^{1{\rm D}}\equiv \frac{1}{2}(a_z-2q_z\cos2t) z^2,\label{Eq:Mo}\ee
that results in a linear equation of motion, with the Mathieu parameter $q_z={2q_5}/({\sqrt{3}\pi})$.
The nondimensional secular frequency of oscillation in the trap (also called the characteristic exponent), $\nu_z(a_z,q_z)$, can be 
approximated in the limit $a_z,q_z^2 \ll 1$ by
\begin{equation}
\nu_z\approx\sqrt{a_z + q_z^2/2}.
\label{eq:omegas}
\end{equation}
The transformation functions between real space coordinates and action-angle variables for Mathieu oscillators can be obtained exactly as presented separately \cite{rfdist}.
Here we use a leading order approximation, obtained using \eq{Eq:TransPseudo}, which for the Mathieu oscillator reduces to
\be \zeta \to z,\qquad \pi_\zeta \to p_z - q_z\sin(2t)z. \label{Eq:TransLin}\ee

The expansion of ${V}_{\rm{ps}}^{1{\rm D}}$ of \eq{2paramSurfacePseudoPot1D} about $\zeta_s$ gives the harmonic oscillator potential,
\be {V}_{\rm{h.o.}}^{1{\rm D}}\equiv \frac{1}{2}\nu_z^2 \zeta^2,\label{Eq:ho}\ee
whose frequency coincides with $\nu_z$ appearing in  \eq{eq:omegas}. 
 The action is related to the energy $E$ by $I\to E/\nu_z$ and the (inverse) action-angle transformation is
\be
\zeta \to \sqrt{\frac{2I}{\nu_z}}\cos\theta, \qquad \pi_\zeta \to-\sqrt{2I\nu_z}\sin\theta,\label{Eq:hotransf}
\ee
where the angle is defined by
\be \theta=\nu_z t+\phi, \ee
with $\phi$ determined by the initial conditions.
The motion lies on an ellipse in phase space with area $J=2\pi I$, evolving clockwise in the $(\zeta,\pi_\zeta)$ plane. 
The approximation of \eq{Eq:TransPseudoPartial} then reduces to
\be\frac{\partial \Lambda_{\rm M.o.}}{\partial p_{z}}\approx \frac{\partial \Lambda_{\rm h.o.}}{\partial p_{z}},\qquad  \frac{\partial^2 \Lambda_{\rm M.o.}}{\partial p_{z}^2} \approx \frac{\partial^2 \Lambda_{\rm h.o.}}{\partial p_{z}^2},\label{Eq:LambdaMaDerivatives}\ee
with the simple formulae for the harmonic oscillator,
\be  \frac{\partial \Lambda_{\rm h.o.}}{\partial p_{z}} = \frac{\pi_\zeta}{\nu_z},\qquad   \frac{\partial^2 \Lambda_{\rm h.o.}}{\partial p_{z}^2} = \frac{1}{\nu_z}.\label{Eq:LambdahoDerivatives}\ee
Throughout the numerical simulations in this work we have compared  the results obtained by using the approximation given in \eq{Eq:LambdaMaDerivatives}, to the exact value of $\partial\Lambda_{\rm M.o.}/\partial p_z$ (obtained using \cite{rfmodes} and presented separately \cite{rfdist}), and the difference is quantitatively very small (given by approximately $q_z^2$).

\section{Analytic Limits of Stochastic Motion in 1D}\label{Sec:Stochastic1D}

In this section we consider the form of the action drift and diffusion coefficients  that enter the FP equation describing different heating and cooling processes in 1D. We summarize here all of the analytic results that apply to 1D motion, with a simplified notation. In some of the cases, the averaging of the FP coefficients can be carried out explicitly and closed form expressions are presented in the following. 
We postpone however a detailed discussion of the results to the following sections (to \seq{Sec:Numerics} where all figures are presented together and the different dynamics compared, and to \seq{Sec:Outlook} which contains a summary and an outlook).
For the 1D motion  
the general FP equation  in action [\eq{Eq:FPIt}], takes the simplified form
\begin{align}
\frac{\partial {P}(I,t) }{\partial t} = -\frac{\partial {S}({I},t) }{\partial I}= - \frac{\partial }{\partial I} \cro{\Pi_I {P}} +\frac{1}{2} \frac{\partial^2 }{{\partial I}^2} \cro{\Pi_{II}{P}}.
\label{AvtransformedFP}
\end{align}

We can gain insight into the cooling dynamics by using the drift and diffusion rates at any value of $I$ to define (respectively) a drift timescale $\tau_{\rm drift}(I)$ and a diffusion timescale $\tau_{\rm diffuse}(I)$, and form a nondimensional coefficient that measures the cooling efficiency,
\be \varepsilon(I)\equiv \frac{ \Pi_{I} I }{\Pi_{II}}\propto \frac{\tau_{\rm diffuse}(I)}{\tau_{\rm drift}(I)}.\label{Eq:varepsilon0}\ee
The sign of $\varepsilon$ depends on the sign of the drift coefficient, and is negative when the laser is cooling the ion (in the mean), while if it is positive, the laser effectively heats the ion.
A large value of $|\varepsilon|$ signifies that the drift time is much shorter than the time it takes to diffuse a similar range of action, so that for $\varepsilon \ll -1$ the cooling is efficient. In contrast, for $-1\lesssim \varepsilon \lesssim 1$ the width of the  ion's distribution in $I$ grows faster than its mean drifts. Diffusive motion has no directionality, and  may equally well lead the ion down or up in action.

For completeness we note that a steady-state of the \eq{AvtransformedFP} [assuming one exists], can be obtained in terms of the FP coefficients, with a distribution which can be thermal-like (exponential) or very different, as will be discussed in a future publication.

\subsection{White Noise Heating in 1D}\label{Sec:Heating1D}

For an ion subject to position-independent Gaussian white noise, the 1D action drift and diffusion coefficients are
\begin{align}
&\Pi_{I}^{\rm w}(I) =D \overline{ \frac{\partial^2 \Lambda}{\partial p_z^2}},
&&\Pi_{II}^{\rm w}(I) =2D \overline{\pare{\frac{\partial\Lambda}{\partial p_z} }^2},\label{Eq:FPaCoeff}
\end{align}
with $D$ being the nondimensional diffusion coefficient. 
White noise can be thought of as being produced by a reservoir at infinite temperature and causes only heating. In this case there is no steady-state distribution.

For the  harmonic oscillator ${V}^{\rm 1D}_{\rm{h.o.}}$ of \eq{Eq:ho}, we have 
\be \Pi_{I}^{\rm w}\to {D}/{\nu_z},\qquad
\Pi_{II}^{\rm w} \to 2DI/\nu_z,\label{Eq:Piaho}\ee
and for the quadrupole potential ${V}_{\rm{M.o.}}^{1{\rm D}}$, the Mathieu oscillator of \eq{Eq:Mo}, within the approximation of \eqs{Eq:TransPseudo}-\eqref{Eq:TransPseudoPartial}, the result is identical to the harmonic oscillator, neglecting as discussed above, a small correction of approximately $q_z^2$. 
In addition, as we confirm by numerical simulations, due to the form of \eq{Eq:FPaCoeff} and by using \eq{Eq:TransPseudoPartial} explicitly, also for motion
within an anharmonic Paul trap potential, the effect of the micromotion is averaged out and can be neglected.

In the case of heating within a quadrupole potential, a solution by separation of variables, with a reflecting boundary condition at $I=0$, and the initial condition ${P}(I,t=0)=\delta(I)$ (the ion starting at the origin), can be integrated in closed form. As can be verified directly by substitution it results in the  time-dependent distribution 
\begin{align}
{P}(I,t)&=\frac{\nu_z }{Dt}\exp\left\{-\frac{\nu_z}{Dt}I\right\},
\end{align}
with the expected mean torus drift
\begin{equation}
\mean{I(t)}={Dt}/{\nu_z}.
\end{equation}
For the harmonic oscillator (for which energy is conserved), this is equivalent to the mean heating rate in energy $
\langle\dot{E}\rangle=\langle\nu_z \dot{I}\rangle=D$.

In the following section we study quantitatively the laser cooling process in the trap, and we will compare the heating rate values calculated for white noise with the laser-induced cooling rates.

\subsection{Laser cooling in a general 1D potential}

For motion in 1D, the FP coefficients of \eqss{Eq:Pijh}{Eq:Pijkh} for cooling in the zero lifetime limit with the decay occurring at the phase-space point of the absorption, are
\bea &\Pi_{I}^{\rm z} =\Gamma\overline{\rho_s\left[
p_{\rm r} \frac{\partial \Lambda}{\partial p_z} +\frac{1}{2} p_{\rm r}^2 (1+ \mu)   \frac{\partial^2 \Lambda}{\partial p_z^2} \right]},\label{Eq:PiIh1D}\\
&\Pi_{II}^{\rm z} = \Gamma p_r^2(1+ \mu)\overline{\rho_s \left(\frac{\partial\Lambda}{ \partial p_z}\right)^2},
\label{Eq:PiIIh1D}\end{align}
with $\mu\equiv\mu_{zz}$ of \eq{Eq:mu}.
All terms in the right hand side of the equations above are functions of $Z_a=\{z,p_z,t\}$, the phase-position, momentum and time where the absorption occurred, that is averaged on a given torus by the definition in \eq{Eq:TorusAverageDef}.

For the finite lifetime case, \eqss{Eq:FPPci}{Eq:FPPcij} become 
\bea &\Pi_{I}^{\rm f} =\Gamma\overline{\rho\left[
p_{\rm r} \frac{\partial \Lambda}{\partial p_z} +\frac{1}{2} p_{\rm r}^2   \left(   \frac{\partial^2 \Lambda}{\partial p_z^2}+\mu \left\langle\frac{\partial^2 \Lambda}{\partial p_z^2} \right\rangle_\Gamma  \right) \right]},\label{Eq:PiIc1D}\\
&\Pi_{II}^{\rm f} = \Gamma p_r^2\overline{\rho  \left[ \left(\frac{\partial\Lambda}{ \partial p_z}\right)^2 + \mu \left\langle\left(\frac{\partial\Lambda}{ \partial p_z}\right)^2 \right\rangle_\Gamma 
\right] },
\label{Eq:PiIIc1D}\end{align}
with the emission point $Z_e$ averaged by the integration of the waiting time distribution $\langle\cdot\rangle_\Gamma$ defined in \eq{Eq:GammaDecay}. 

The dynamics of the laser cooling around a given torus are determined both by the form of the potential (that depends on the displacement of the ion), and by the velocity, as compared to a relevant scale determined by the laser parameters. The limits of low velocity and of high velocity lend themselves to expansion in a suitable, distinct small parameter. For motion in a quadrupole potential this allows a simplification of the FP coefficients, explored in the following two subsections for 1D motion.

\subsection{The linear limit of laser cooling}\label{Sec:Final1D}

In a Paul trap whose potential reduces to a quadrupole at its center (as in linear Paul traps and surface-electrode traps, but not in higher multipole traps), and in the absence of excess micromotion, the ion velocities can be assumed to be small at the center of the trap. These two conditions define a specific limit of the cooling, that can be analyzed analytically. 
The Lorentzian can be linearized in the velocity, with the coefficients written in a well-known form \cite{javanainen1980,leibfried2003}, for $s\ll 1$,
\be 
p_{\rm r}\Gamma\rho(p_z)\approx F_{\rm r} + \gamma v_z,\label{Eq:LorentzianLin}\ee
with
\be F_{\rm r}=\frac{p_{\rm r}\Gamma s/2}{1 +(2\Delta/\Gamma)^2}, \quad \gamma = \frac{4k p_{\rm r}s \Delta/\Gamma}{\left[1 +(2\Delta/\Gamma)^2\right]^2}.\label{Eq:F0gamma}\ee
Here $F_{\rm r}$ gives a mean radiation force (for ${p}_z=0$), and $\gamma/m$ (with $m=1$ in the rescaled units) is the damping rate. This linearization is valid if
\be k v_z \ll \left[\Gamma^2 +4\Delta^2\right]/(8|\Delta|)\label{Eq:kvLin},\ee
which can be rewritten in terms of the action,
\be I \ll I_{\rm linear} = \left(\frac{\Gamma^2+4\Delta^2}{8k\Delta}\right)^2\frac{1}{2\nu_z}.\label{Eq:Ilinear}\ee
In this approximation, since the Lorentzian is linear in the momentum, and the partial derivatives of the action are (at most) linear in the momentum (for a quadrupole potential), the resulting cooling coefficients are at most linear in the actions. For the harmonic oscillator the integration in closed form of the action drift and diffusion coefficients is straightforward and we can write 
\be \Pi_{I}^{\rm l} = \gamma I+h_z/2,\qquad \Pi_{II}^{\rm l} = {h}_z I,\label{Eq:PiIl1D}\ee
with
\begin{align}
h_z= p_{\rm r}F_{\rm r}(1+\mu) /\nu_z. \label{Eq:FP1Dh}
\end{align}
Then the linear limit of the cooling is characterized by a thermal equilibrium-like distribution (for $\Delta<0$), determined by the balance of momentum dissipation and diffusive heating. The mean value and standard deviation of the action, $\langle I\rangle =\sqrt{\langle (I-\langle I\rangle)^2\rangle }= I_{\rm limit}$, reads 
\be I_{\rm limit} \equiv  \hbar \frac{\Gamma}{8 \nu_z} ( 1 + \mu) \left[ \frac{\Gamma}{2 \left|\Delta\right|} + \frac{2 \left|\Delta\right|}{\Gamma}\right].\label{Eq:Ilimit} \ee

The expression above is identical to the known results for the thermal distribution of an ion cooled within a 1D harmonic oscillator \cite{leibfried2003}, in the limit of $s\ll 1$. 
Remarkably we find in the numerical simulations presented in \seq{Sec:FinalNumerics} that the results for the Mathieu oscillator are nearly identical [up to the accuracy of \eq{Eq:LambdaMaDerivatives}, neglecting a correction of approximately $q_z^2$].
Analytic expressions for the coefficients can be obtained for general 3D motion in a Mathieu oscillator potential, which is beyond the scope of the current work and will be presented separately. 

\subsection{The (quadrupole) fast particle limit of laser cooling}\label{Sec:DopplerFast}

A second regime of motion that allows a simplification of the expressions of the action coefficients of laser cooling is the (quadrupole) fast particle limit, where the motion is of a large enough amplitude, within a quadrupole potential. In this case, the ion velocity creates a large Doppler shift which tunes the cooling light {\it out} of the resonance with the ion for most of the time. In a harmonic motion  the measure on the invariant torus where the light field is resonant with the ion, bounded within the strip $\vec{k}\cdot\vec{v} = v_0\pm \delta v$ [the resonance has width $\delta v$ around $v_0$, defined in \eq{Eq:Lorentzian}], is independent of the action, and its fraction of the total torus measure decreases with $I$. Within the tori averages, the contribution from the resonance region decreases accordingly, while the contribution from the rest of the torus increases with $I$ due to the tails of the Lorentzian distribution, which eventually dominate the torus averages.

The cooling and diffusion terms within the fast particle limit
are derived in \app{Sec:DopplerFastApp} on the basis of an expansion for the
Lorentzian expression of the excited state occupation. For the case that the laser propagates along one of the principal axes of the motion [here, $\vec{k}=k\hat{z}$], and if in addition the motion is descirbed by a harmonic oscillator [\eq{Eq:hotransf}], the fast particle limit  expressed in terms of the action is 
\be I \gg I_{\rm fast} \equiv  \frac{\max\left\{4\Delta^2,\Gamma^2/4\right\}}{\nu_z k^2/2}.\label{Eq:DeltaFast2}\ee
We have  integrated the drift and diffusion coefficients in closed form [\eqss{Eq:PiIfho}{Eq:PiIIfho}], and a basic property of the coefficients in this limit is their action dependence \cite{javanainen1981laser},
\be \Pi_I^{\rm f.p.} \propto -1/\sqrt{I},\qquad \Pi_I^{\rm f.p.} \propto \sqrt{I}.\label{Eq:PiIfMo}\ee 
 We note that the zero lifetime treatment for a harmonic oscillator gives the same cooling rate in terms of the energy (which has been used, e.g., in \cite{wesenberg2007,PhysRevA.96.012519}), but results in a constant (independent of $I$) diffusion term [\eq{Eq:PiIIhho}], which is incorrect since \eq{Eq:PiIfMo} shows a scaling $\propto{\sqrt{I}}$.

As discussed in \seq{Sec:CoolingPaul}, for motion within a Mathieu oscillator potential we find by numerical integration that the action dependence is identical in form [obeying \eq{Eq:PiIfMo}], with the prefactors depending on the Mathieu oscillator parameters.
The implications of these functional relations are discussed in \seq{Sec:CoolingPaul}, and the deviations from them for motion in the anharmonic rf potential, in \seq{Sec:CoolingNonlinear}. 

\section{A Study of Heating and Cooling in 1D}\label{Sec:Numerics}

In this section we present a detailed numerical study of heating and cooling processes for ion motion in 1D. We compare the results for motion within the four realizations of a trap potential described in \seq{Sec:Hamiltonian1D}. 
For reference, these are summarized in Table \ref{table:potentials}, together with their acronyms (used in the figures and equations).
In Table \ref{table:Pis} we summarize the notation indicating the action drift and diffusion coefficients that correspond to the different physical processes and parameter regimes, presented in the figures that follow in this section and the equations used to calculate them..

\begin{table}
\centering
 \begin{tabular}{|c | c | c|} 
 \hline
 Potential is & Time-independent & Time-dependent \\  
 \hline\hline 
 Quadrupole & \,Harmonic oscillator\, &\, Mathieu oscillator  \,
\\ & (h.o.) & (M.o.)
 \\ 
 \hline
 Anharmonic  & Pseudopotential   & Rf potential\\
 & (ps)  & (rf) \\
 \hline \hline 
\end{tabular}
\caption{The four different types of trap potentials, summarized with their acronyms used in the figures to follow and in the equations. Depending on the Paul trap type, along each of its spatial dimensions the potential can be quadrupole or anharmonic, and either static or periodically driven.}

\label{table:potentials}
\end{table}
\begin{table}
\centering
 \begin{tabular}{|c | c|} 
 \hline
 Notation &  Definition \\  
 \hline\hline 
 $\Pi_I^{\rm w},\,\Pi_{II}^{\rm w}$ & \,\,Position-independent additive {\underline w}hite noise\,\,  
\\ 
 & \eq{Eq:FPaCoeff}  \\ 
 \hline
 $\Pi_I^{\rm z},\,\Pi_{II}^{\rm z}$ & Laser cooling in the {\underline z}ero lifetime  
\\  &  (heavy particle) limit\, \eqss{Eq:PiIh1D}{Eq:PiIIh1D}\\
 \hline
 $\Pi_I^{\rm f},\,\Pi_{II}^{\rm f}$ & Laser cooling in the {\underline f}inite lifetime limit 
\\  & \eqss{Eq:PiIc1D}{Eq:PiIIc1D} \\
 \hline
 $\Pi_I^{\rm f.p.},\,\Pi_{II}^{\rm f.p.}$ & Laser cooling in the {\underline f}ast {\underline p}article limit
\\  & \eq{Eq:PiIfMo} \\
 \hline
 $\Pi_I^{\rm l},\,\Pi_{II}^{\rm l}$ & Laser cooling in the {\underline l}inear limit 
\\  & \eq{Eq:PiIl1D} \\
 \hline \hline 
\end{tabular}
\caption{The different types of action drift and diffusion coefficients, used in figures to follow and in the equations.}
\label{table:Pis}
\end{table}

For most of the calculations in the following (except where explicitly stated otherwise), we set the laser detuning to the Doppler detuning, with a low saturation parameter,
 \be  \Delta=-\Gamma/2, \qquad s=0.01. \label{Eq:LaserParams2}\ee 
For the chosen $^9{\rm Be}^+$ ion and in the nondimensional units introduced in \eq{Eq:rescaling}, we have $\Gamma = 0.38$,  see \app{Sec:Nondim} for details of the laser (and trap) parameters. Motional heating is represented by a white noise rate, and we choose a representative value of the nondimensional diffusion coefficient $D$ that corresponds to a heating rate in units of motional quanta per second near the effective potential minimum (see \app{Sec:Nondim}), of 
\be \dot{\tilde{n}}=0.1\,{\rm ms}^{-1},\label{Eq:dotn}\ee
except in \fig{fig:Heavy}.

\subsection{Cooling in the low velocity regime}\label{Sec:FinalNumerics}
  
 \begin{figure}[!t]
\includegraphics[width=3.3in]{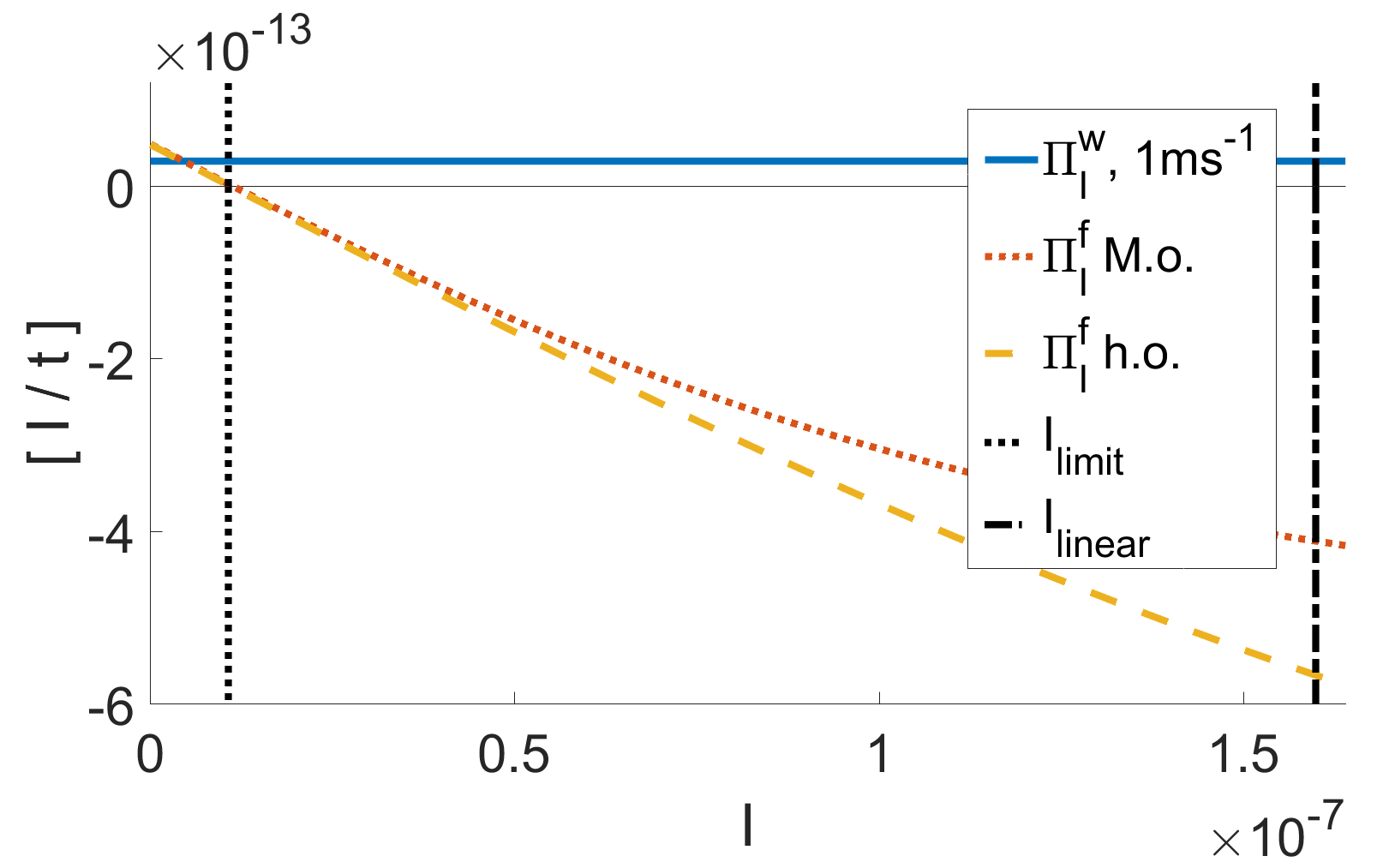}
\caption{The action drift coefficients for laser cooling [$\Pi_I^{\rm f}$, \eq{Eq:PiIc1D}], and for white noise heating [$\Pi_I^{\rm w}$, \eq{Eq:FPaCoeff}] in nondimensional units, for low amplitude motion. In this regime the rf potential is accurately approximated by its Mathieu oscillator limit. The trap parameters, laser parameters, and heating rate parameters are given in \app{Sec:Nondim}. For this figure only, we have set the saturation parameter to $s=0.001$, and the heating rate, in terms of quanta per unit time, to $\dot{\tilde{n}}=1\,$ms$^{-1}$. 
The vertical lines show $I_{\rm linear}$, the border of validity of the linear limit of the cooling [\eq{Eq:Ilinear}], and the action at the Doppler cooling limit, $I_{\rm limit}$ [\eq{Eq:Ilimit}]. It can be seen  how the Mathieu and the harmonic oscillator (of the same secular frequency)  converge for $I\ll I_{\rm linear}$ to the same curve, and for both curves the drift vanishes as $I=I_{\rm limit}$.  See the text for a discussion of the competing effects of heating $\Pi_I^{\rm w}\propto D$ and cooling $\Pi_I^{\rm f}\propto s$.}
\label{fig:Heavy}
\end{figure}

Although ions are cooled from high-amplitude motion towards low amplitude, we present our detailed study of laser cooling starting with low amplitude motion with $I\lesssim I_{\rm linear}$ [defined in \eq{Eq:Ilinear}]. 
 The action drift coefficient $\Pi_I^{\rm f}$ is composed of the sum of the absorption term (linear in $p_{\rm r}$), and the emission term ($\propto p_{\rm r}^2$). The dynamics is determined by the competition of the two terms, since the first is negative (for $\Delta<0$) and the second is positive, resulting from the diffusion in phase-space. The action value for which $\Pi_I^{\rm f}=0$ (and has a negative slope), is where the mean drift balances, with the ion being heated (by momentum diffusion) for lower values of $I$, and cooled back from higher $I$. Figure \ref{fig:Heavy} shows that for $I\ll I_{\rm linear}$, both the Mathieu and harmonic oscillator drift rates converge to the same curve, reducing in this limit to the expression of \eq{Eq:PiIl1D}. 
 The numerically calculated zero-crossing for both the harmonic oscillator and the Mathieu oscillator  coincide very closely with the (thermal-equilibrium-like) limit of \eq{Eq:Ilimit}. The crossing points are independent of $s$ (in the limit $s\ll 1$) but the slope is proportional to $s$, which is important if heating of a comparable rate is present. The action drift rate resulting from white noise heating at a rate of 1 quantum per millisecond (10 times more than in the following figures, a value that can be considered as high but not excessive in current state-of-the-art traps), is shown for comparison. Such a high heating rate will shift the cooling limit to a action value where the sum $\Pi_I^{\rm w}+\Pi_I^{\rm f}$ crosses 0, and increase the final mean value of the action (with these parameters to $\sim 2I_{\rm limit}$). However, the saturation parameter in this figure is taken to be $s=0.001$, much lower than what is typically used in experiments.  Since the heating drift rate scales linearly with $D\propto\dot{n}$, and the cooling rate scales linearly with the intensity $\propto s$ (for $s\ll 1$), the cooling limit is quite insensitive to heating at this order of magnitude.
For $I\gtrsim 0.5\times 10^{-7}$, i.e.~where the condition $I\ll I_{\rm linear}$ no longer holds, the Mathieu and harmonic oscillator cooling rates begin to deviate (with the former being smaller).

 \subsection{Cooling in the high velocity regime of a quadrupole potential}\label{Sec:CoolingPaul}

\begin{figure}[!t]
\includegraphics[width=3.3in]{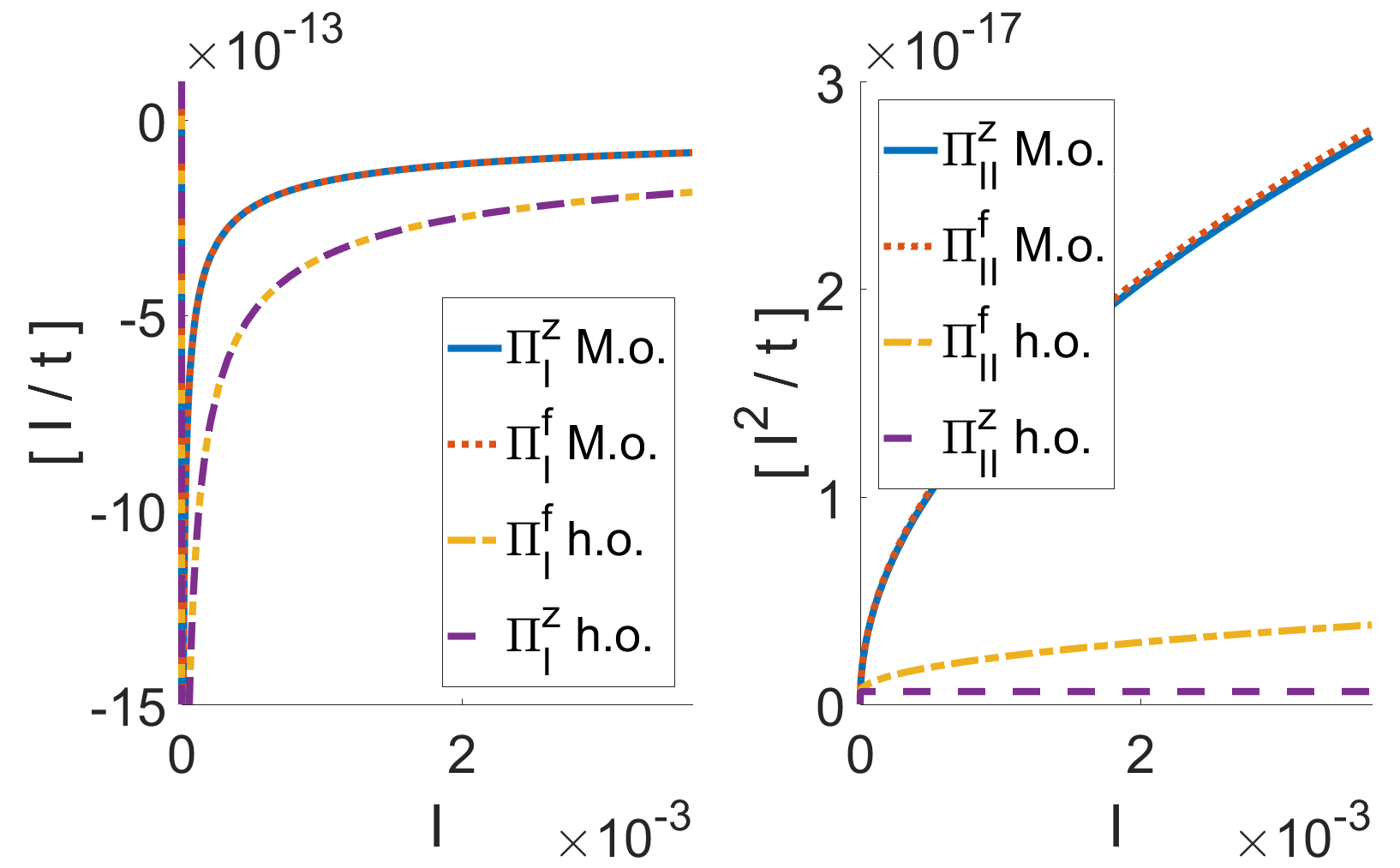}
\caption{(a) The action drift and (b) diffusion coefficients in the large amplitude regime within a quadrupole potential, calculated using the expressions of the zero lifetime limit ($\Pi_I^{\rm z}, \Pi_{II}^{\rm z}$) and the finite lifetime treatment  ($\Pi_I^{\rm f}, \Pi_{II}^{\rm f}$), with trap and laser parameters given in \app{Sec:Nondim} and \eq{Eq:LaserParams2}. (a) The drift curves calculated  (for the harmonic oscillator and separately for the Mathieu oscillator) using the zero lifetime expressions very closely coincide with their values within the finite lifetime treatment. (b) For the harmonic oscillator, the diffusion rate in the zero lifetime limit saturates at a constant value, which quantitatively and qualitatively differs from the results of the finite lifetime treatment. For the Mathieu oscillator the diffusion coefficient calculated using the zero lifetime expression ($\Pi_{II}^{\rm z}$) nearly coincides with the curve for the finite lifetime ($\Pi_{II}^{\rm f}$). See the text for a detailed discussion.}
\label{fig:hvsc}
\end{figure}

The Mathieu and harmonic oscillator cooling rates, although deviating from each other for $I\gtrsim I_{\rm linear}$,  both show 
an asymptotic slow approach towards zero at high action.
The numerically calculated values of the FP coefficients are shown in \fig{fig:hvsc} for motion within a purely quadrupole potential up to a large   amplitude.
 Although the zero lifetime treatment does not apply to the parameters of our study, we find it instructive to compare the coefficients calculated by using the expressions for zero lifetime with the finite lifetime treatment. 
 For harmonic oscillator motion, the action drift (cooling) rates $\Pi_I^{\rm z}$ and $\Pi_I^{\rm f}$ nearly coincide in this limit and are given by $\Pi_I^{\rm f.p.}\propto -1/\sqrt{I}$ of  \eq{Eq:PiIfho}, since the correction to the drift coefficient that comes from the spontaneous emission term $\propto \mu$ in \eq{Eq:PiIc1D}, where the zero lifetime and finite lifetime expressions differ, is negligible for these parameters.  The diffusion coefficient calculated by $\Pi_{II}^{\rm z}$ for a harmonic oscillator can be seen to saturate [\eq{Eq:PiIIhho}], while it in fact grows as the amplitude of the motion gets larger [$\Pi_{II}^{\rm f.p.}\propto \sqrt{I}$, \eq{Eq:PiIIfho}].

For the time-dependent Mathieu oscillator, the action drift rates again nearly coincide within the zero and finite lifetime limits. The diffusion coefficients show a more distinct behavior. The Mathieu oscillator diffusion is much larger than the harmonic oscillator diffusion. In addition, in contrast to the harmonic oscillator case, the large diffusion is predicted by the zero lifetime expressions. This is because the emission occurring at some later time after the absorption [as expressed by the averaging over the decay process, $\langle\cdot\rangle_\Gamma$ in \eq{Eq:PiIIc1D}], is no longer the main cause for the large diffusion. Rather, due to the fast micromotion, the absorption and following emission occur at various phase-space points along the trajectory, where the kick to the action (determined by $\partial\Lambda/\partial p_z\propto \pi_\zeta$) can be both positive and negative. Although the average of all kicks remains negative, their variance is much bigger.

 As \fig{fig:hvsc} suggests, although the dynamics within the time-independent potential are simpler than within the time-dependent one, laser cooling with the latter can in some cases be well described using the simpler zero lifetime limit. Formulating a criterion that explicitly states under which circumstances the tori averages of the drift and (in particular) the diffusion coefficients for the finite lifetime limit can be approximated by the zero lifetime limit appears to be hard. However, since the zero lifetime expressions are simpler and faster to calculate numerically, it is worth having this possible shortcut in mind when performing a numerical study of laser cooling based on concrete trap and laser parameters.

Our results allow us to derive a further important conclusion regarding the nature of the cooling. The numerically calculated coefficients show that in the fast particle limit, the asymptotic behaviour within the Mathieu oscillator has the same functional dependence as that within a harmonic oscillator, $\Pi_{I}^{\rm f.p.}\propto -1/\sqrt{I}$ and $\Pi_{II}^{\rm f.p.}\propto \sqrt{I}$, which  immediately implies that in this limit, the efficiency coefficient of \eq{Eq:varepsilon0} is independent of the amplitude, since
\be \varepsilon^{\rm f.p.}(I)\equiv \frac{ \Pi_{I}^{\rm f.p.} I }{\Pi_{II}^{\rm f.p.}}={\rm const}<0.\label{Eq:epsilonf}\ee
We note that in this ratio, the saturation parameter drops out (in our low saturation limit). As long as an appropriate choice of the  laser parameters guarantees that $|\epsilon^{\rm f.p.}(I)|\gg 1$, the cooling process is efficient (and non-diffusive), independent of the action. 
  
 \subsection{Cooling in an anharmonic Paul trap potential}\label{Sec:CoolingNonlinear}
 
\begin{figure}[!t]
\includegraphics[width=3.3in]{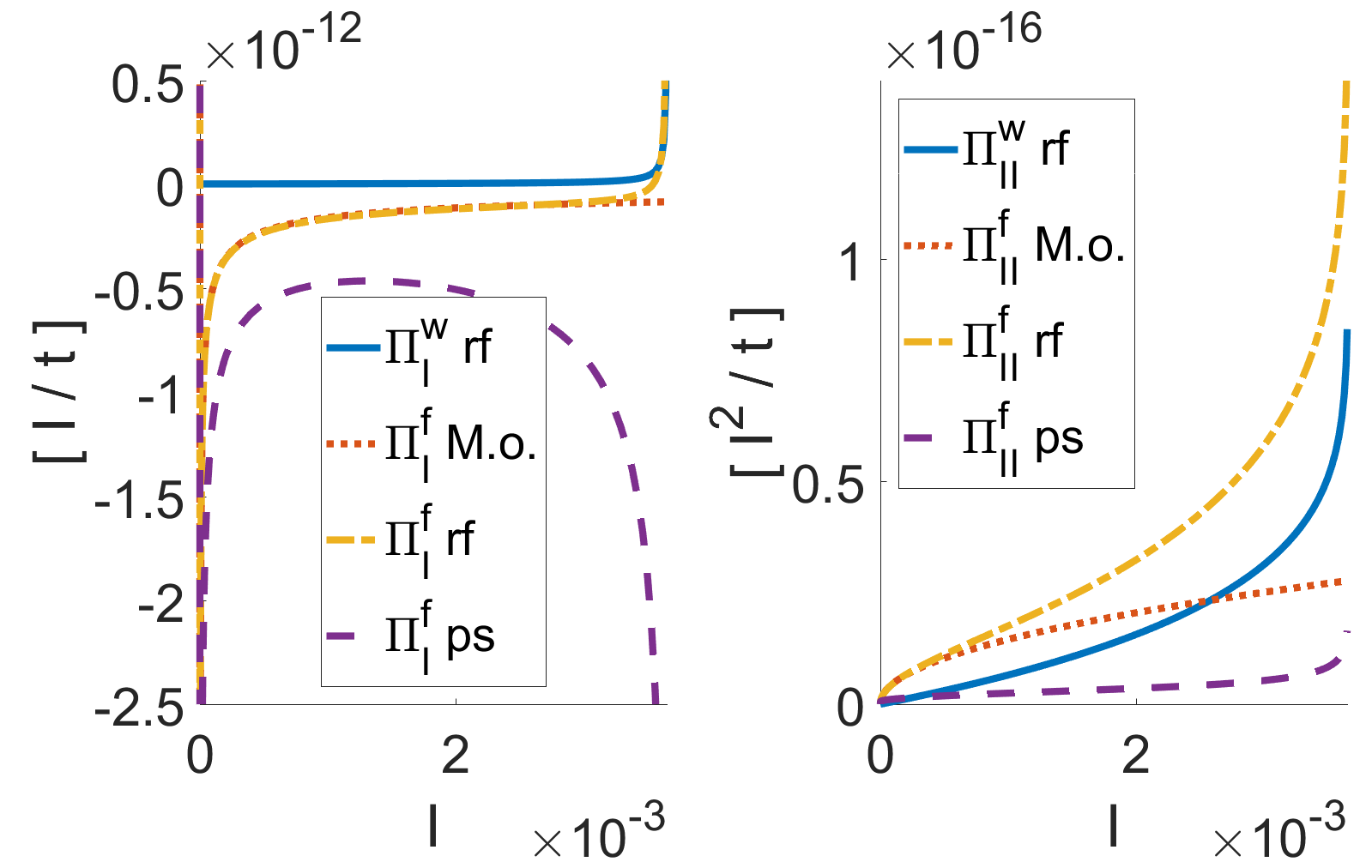}
\caption{ (a) The action drift and (b) diffusion coefficients for white noise heating and laser cooling within different types of Paul trap potentials. The laser cooling coefficients ($\Pi_I^{\rm f}$ for the drift and $\Pi_{II}^{\rm f}$ for the diffusion) are compared for motion within a Mathieu oscillator, the surface trap full rf potential, and its pseudopotential approximation. The coefficients of white noise heating ($\Pi_I^{\rm w}$ and $\Pi_{II}^{\rm w}$, within the rf potential) are shown as well. The trap and laser parameters are as in \fig{fig:hvsc}, and the heating rate is given in \eq{Eq:dotn}. We note that the $I$ axis here extends to  $I=3.65\times 10^{-3}$ (where $\nu(I)\approx 0.03$), at the border where the rf potential motion becomes chaotic for the presented parameters). See text for details and a discussion. 
}\label{fig:Fast}
\end{figure}

 Comparing the cooling coefficients for motion within the surface-electrode trap potential, \fig{fig:Fast}(a) shows that a calculation using the pseudopotential results in a laser-induced drift rate very different from the rf potential. Cooling within the rf potential is well described by the Mathieu oscillator approximation throughout most of the trap. However, beyond a certain amplitude of motion, the cooling turns into heating as evidenced by the drift coefficient becoming positive. In this region, the corrections to the cooling rate coming from the two terms at order $p_{\rm r}^2$, dominate the drift rate.  The heating from  white noise corresponding to a heating rate of $0.1\,{\rm ms}^{-1}$ [\eq{Eq:dotn}] is shown for comparison, calculated for the motion in the anharmonic potential. Both can be seen to start diverging in the region of motion that approaches the separatrix. This results from the term $\propto \nu^{-3}dI/d\nu$ that enters $\Pi_I^{\rm f}$ through $\partial^2\Lambda/\partial p_z^2$ [\eq{Eq:TransPseudoPartial}], and becomes important only close enough to the separatrix, although it should be noted that $\nu(I)$ at the maximal value of $I$ in \fig{fig:Fast} is still $\sim 1/4$ of its value at the trap center, so our adiabatic approximation still holds. The action axis extends up to the maximal  value for which a simulation of the full rf potential shows that the ion is still bounded (by the last unbroken torus \cite{rfchaos}). For the presented parameters the voltage on the rf electrodes is within the border of validity of the pseudopotential, and the chaotic region close to the separatrix is very small.

\begin{figure}[!t]
\includegraphics[width=3.3in]{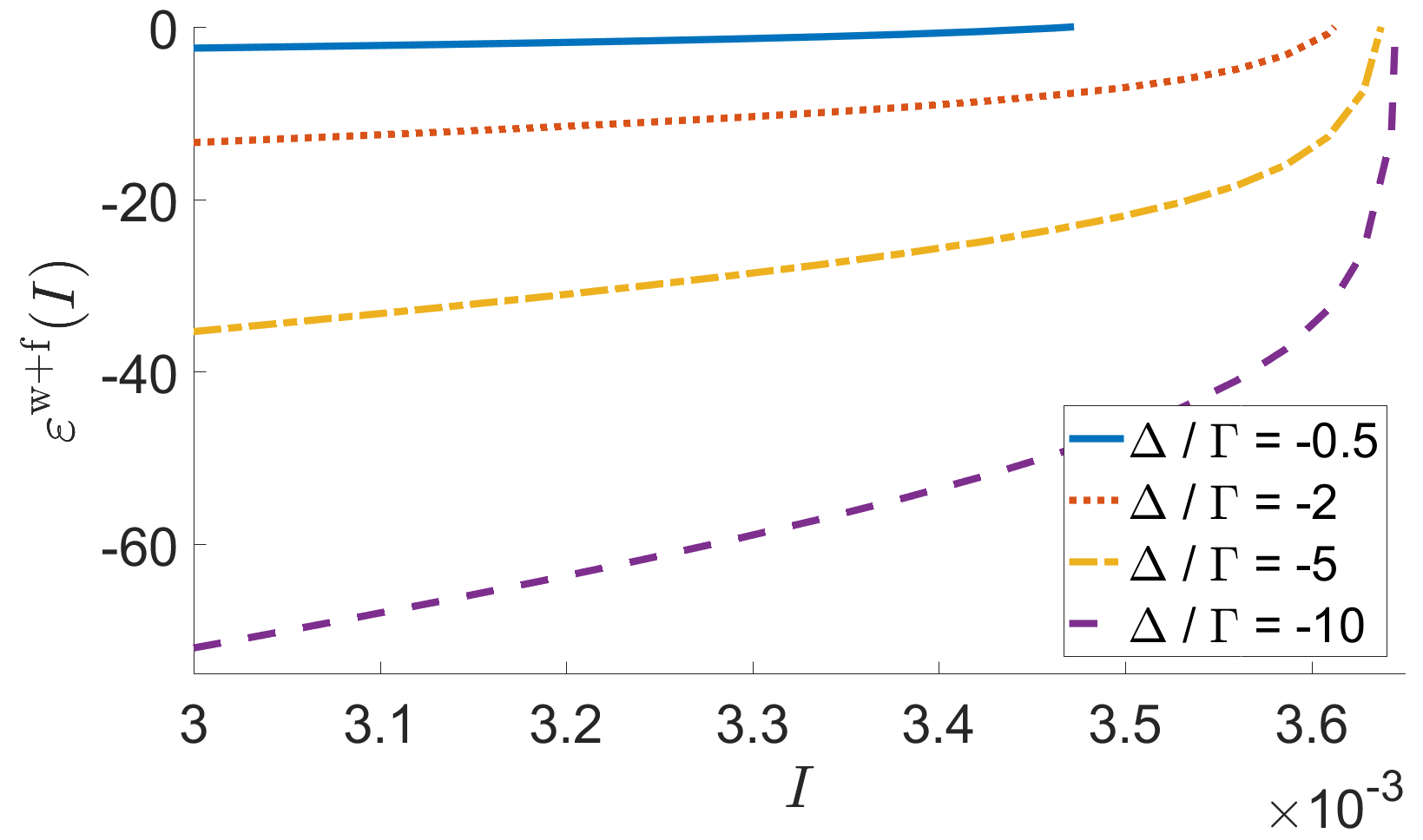}
\caption{The cooling efficiency coefficient $\varepsilon^{\rm w+f}(I)$ of \eq{Eq:epsilonac}, in the large amplitude regime of a surface-electrode trap. Four values of of the detuning $\Delta$ are shown, with the rest of the parameters as in  \fig{fig:Fast}. The vertical axis is truncated at $\varepsilon=0$, beyond which the combined effect of the noise and laser is a mean positive drift. However, already in the range $\varepsilon\gtrsim -1$, the cooling is inefficient, with the diffusion (due to the noise and  laser together) dominating the drift, which can quickly lead the ion above the trap's barrier.
}\label{fig:epsilon}
\end{figure}

The trap anharmonicity plays a bigger role in the diffusion coefficients [\fig{fig:Fast}(b)]. Here again, the pseudopotential curve is quantitatively very different. Also the approximation of the rf potential by a Mathieu oscillator, results in an underestimate of the action diffusion for nonlinear motion. Turning to the cooling efficiency 
\be \varepsilon^{\rm f}(I)=\frac{ \Pi_{I}^{\rm f} I }{\Pi_{II}^{\rm f}},\label{Eq:varepsilon}\ee
 for low detuning as in \fig{fig:Fast} (where $\Delta=-\Gamma/2$) we find that  $\varepsilon^{\rm f}(I) \ll -1$ throughout most of the trap (beyond the very low amplitude motion of thermal equilibrium), and it increases only for very high amplitude motion ($\varepsilon(I) > -1$ for $I\gtrsim 3.5\times 10^{-3}$). This border is close to (but still lower than) the point where the laser would start heating the ion ($\varepsilon(I)>0$). The reason that $|\varepsilon(I)|$ becomes of order 1 is inherent to the anharmonic rf potential. As can be seen in  \fig{fig:Fast}, the diffusion grows more steeply than in a quadrupole dependence (for which $\Pi_{II}^{\rm f.p.}\propto \sqrt{I}$), already at values of $I$ where the drift is still close to its quadrupole behaviour. We can conclude that due to the nonlinearity the cooling efficiency strongly decreases with the amplitude, and beyond a certain threshold action, the ion motion under cooling becomes diffusive in nature.
 
 Nonetheless, this threshold can be pushed up by varying other parameters of the cooling [though not the intensity, which cancels out in \eq{Eq:varepsilon}]. We find that for a larger detuning the cooling efficiency can be increased. This is clear from, e.g.~\eq{Eq:PiIfho}, for a quadrupole potential. The action value above which $\varepsilon(I)\gtrsim -1$ (where the motion under cooling becomes diffusive), depends on other parameters of the trap, laser, and on their combined effect together with the white noise. The white noise diffusion coefficient is plotted in \fig{fig:Fast}(b), for comparison with the laser induced diffusion. For the chosen parameters we see that the laser diffusion is comparable to the noise diffusion for high amplitude motion. The linearity of the FP equation allows us to examine the cooling efficiency in the presence of white noise,
\be \varepsilon^{\rm w+f}(I)=\frac{ \left(\Pi_{I}^{\rm w}+\Pi_{I}^{\rm f}\right) I }{\Pi_{II}^{\rm w}+\Pi_{II}^{\rm f}}, \label{Eq:epsilonac} \ee
presented in \fig{fig:epsilon}.
For the lowest value of the detuning, the cooling becomes inefficient already at $I\gtrsim 3.36\times 10^{-3}$, due to the white noise (however increasing the laser intensity reduces the relative importantce of the noise contribution). For increased detuning, the limits of this region can be pushed noticeably up. A detailed study of this regime of high amplitude motion could prove important for optimizing ion loading, and we will examine some aspects of ion dynamics subject to a large laser detuning in \cite{rfcycles}.

\section{Summary}\label{Sec:Summary}

The main purpose of the current paper has been to lay down a framework  for treating stochastic processes in rf traps, throughout the regular parts of the unperturbed Hamiltonian phase-space. In general this requires accounting for the trap's periodic drive and its anharmonicity. This can be achieved by employing action-angle coordinates, which also permit  significant simplification of the treatment of slow stochastic processes by integrating over the angles. 
We have kept the derivations of the theory completely general for 3D motion within these assumptions, which  should allow extending our detailed analytic and numerical study for 1D motion, to more spatial dimensions and even to more ions. We begin this section with a summary of the main results.

In \seq{Sec:Heating1D} we study heating by additive, position-independent Gaussian white noise (modelling fluctuating electric fields). 
The simplest case is that of a quadrupole potential, where the ion is heated up at a constant rate, which is approximately equal for both the harmonic and the Mathieu oscillator.
Studying an anharmonic potential, we find that the drift and diffusion rates significantly increase on the high action tori of the trap and, depending on the parameters, may become comparable with the effects of laser cooling. 

 \begin{figure}[!t]
\includegraphics[width=3.3in]{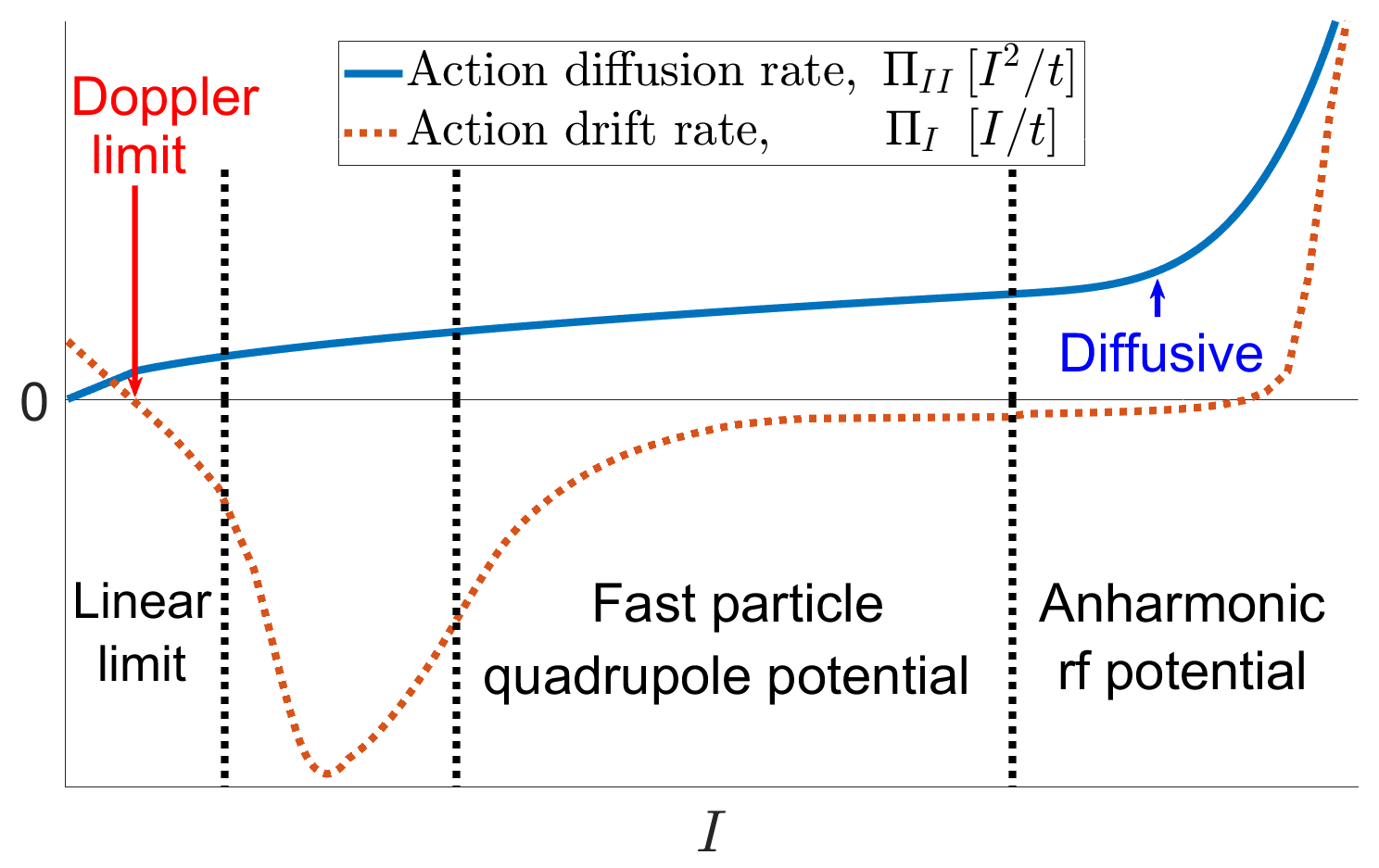}
\caption{A schematic depiction of the drift and diffusion rates of the action $I$, for an ion being  laser-cooled in the different regimes of motion within  a  realistic surface-electrode Paul trap with substantial anharmonicity far away from the effective potential minimum. The axes are not to scale, and the two plotted quantities have different dimensions -- see the legend. In the linear limit of the cooling (in the limit of low velocity without ``excess micromotion''), the cooling drift and diffusion are linear in the action, and converge to those found for a time-independent trap, with the drift coefficient crossing zero at the action corresponding to the Doppler cooling limit. For high velocity and still within an approximately quadrupole potential, the cooling rate drops as $\propto 1/\sqrt{I}$ and the diffusion rate grows as $\propto \sqrt{I}$. For a very high amplitude of motion within the anharmonic, rf potential, the diffusion grows more sharply and may dominate the drift (so the distribution broadens faster than its mean is cooling down), and the drift itself may become positive (turning into effective heating). In this region the ion is likely to escape the trap, however both effects can be partly remedied by using a large detuning of the cooling beam.}
\label{fig:Schematic}
\end{figure}

For laser cooling dynamics, in addition to the degree of the anharmonicity of the potential (that varies with the ion's displacement from the trap center), a velocity scale determined by the laser parameters is important. When the ion's velocity is small enough the Lorentzian describing the absorption probability can be linearized in the velocity, while when the ion's velocity is large enough the ion spends most of its time on the torus within the tails of the Lorentzian.
The different regimes of laser cooling are depicted schematically in \fig{fig:Schematic}. For low velocity the cooling coefficients become linear in the action and we find (in \seq{Sec:Final1D} and in \seq{Sec:FinalNumerics}) that they nearly coincide for a Mathieu oscillator in this limit with those of the corresponding harmonic oscillator.

For a quantitative measure of the cooling efficiency we define a nondimensional coefficient $\varepsilon(I)$ [in \eq{Eq:varepsilon0}] proportional to the diffusion timescale divided by the drift timescale. A large negative value of $\varepsilon(I)$ indicates that the ion drifts towards lower action at a rate which overcomes the spread of its probability distribution.
For high velocity motion within a quadrupole potential
we find (\seq{Sec:CoolingPaul}) that the cooling efficiency parameter is independent of the action, $\varepsilon(I)={\rm const}$, so the cooling remains effective for any amplitude, if the parameters are chosen to guarantee $\varepsilon\ll -1$.

This simple picture breaks down, however, when the anharmonicity of the potential can no longer be  neglected.
We find (\seq{Sec:CoolingNonlinear}) that for a typical low detuning value (optimal for reaching the lowest cooling limit around the trap center), the  trap's anharmonicity in combination with the micromotion lead to diffusive dynamics (dominated by nondirectional diffusion), with $\varepsilon(I)\gtrsim -1$ as the motion amplitude increases within a surface-electrode trap. Moreover, in the very highly anharmonic region close to a separatrix, the drift rate may become positive, with the laser effectively heating the ion past the trap's boundaries. We also find that a laser detuning much larger than the Doppler detuning allows cooling of the ion from much higher action values. 
 
We can draw general conclusions about the usefulness of the time-independent pseudopotential approximation. As discussed in \cite{rfchaos}, 
there exists a regime of parameters where an ion's motion in a Paul trap is nearly integrable, whence the structure of the phase-space  can be well approximated by the pseudopotential, which is simpler to tackle theoretically and to simulate numerically. As we find here, the pseudopotential is also sufficient for a quantitative calculation of white noise heating, where micromotion can be neglected. 
In contrast, except in its linear limit, laser cooling requires that the micromotion be accounted for even for a quadrupole potential. Moreover, neglecting the micromotion in anharmonic regions within a surface-electrode trap leads to incorrect descriptions of the dynamics. 
With the (realistic) values of $q_z$ and $\nu_z$ stated in \eqss{Eq:TrapParamsFP1}{Eq:TrapParamsFP2}, the pseudopotential approximation fails even for a small micromotion amplitude. 

Finally, we note that as discussed in \seq{Sec:Hamiltonian}, we have used an approximate  canonical transformation to obtain the rf potential phase-space variables from the pseudopotential ones, simplifying the numerical analysis significantly in the 1D case. This approach neglects corrections of order $\nu(I)^2$, and, as we have verified with a Mathieu oscillator (for which exact analytic expressions are available \cite{rfmodes}), amounts to roughly a few percent for our parameters. The frequency $\nu(I)$ and with it the expected inaccuracy only decrease with the amplitude of motion due to anharmonicity in the potential studied here [see \fig{fig:ps}(b)], provided that the motion remains nearly integrable. Close enough to a separatrix of the pseudopotential the micromotion introduces chaotic dynamics, posing a natural boundary to the applicability of the presented theory. For the presented parameters the chaos is limited to a very small region of action near the separatrix.

\section{Outlook}\label{Sec:Outlook}

 For a study of the stochastic processes of a single ion, the theory can be directly applied in regions of regular motion in those cases where the actions can be calculated. For a quadrupole potential there is no chaotic motion and moreover analytic expressions can be used. If the potential attains a weak anharmonic component, it may be treated perturbatively starting from a quadrupole potential up to some amplitude scale, which makes the  calculation feasible \cite{PhysRevE.67.061111}. 
Even for potentials where the anharmonicity is relatively strong (requiring a nonpertubative treatment), there are important cases with a symmetry axis along which the motion nearly decouples from the radial plane \cite{wilson2014tunable,rfchaos}. The pseudopotential phase-space for motion in the two radial coordinates is 4D and amenable to an analysis using 2D planar Poincar\'e surfaces of section, which allow one to calculate the actions. The drift and diffusion coefficients become functions of two variables, which can be readily visualized and analyzed and the micromotion can be accounted for using the canonical transformation employed here. Numerically, one complication in such a study (beyond the tools that have been used in this work and in \cite{rfchaos}), could arise from the need to obtain smooth enough maps of  phase-space allowing one to take partial derivatives of the actions. 

In this work we have focused on a study of the drift and diffusion coefficients and the information that can be extracted directly from them. With different initial conditions, the drift and diffusion coefficients can be used to obtain time-dependent solutions of the FP equation, or to obtain some partial statistics such as the mean time to escape the trap in the absence of cooling, or in contrast, to be cooled to the cooling limit from high amplitude. Our theory can be directly applied to the analysis of Doppler cooling thermometry \cite{wesenberg2007,PhysRevA.96.012519,laupretre2018controlling,meir2018direct} and related methods \cite{clark2010detection}.
A complex  setup can be treated by adding the drift and diffusion coefficients calculated separately for each stochastic process. 
 To account for a spatially inhomogeneous laser profile, the saturation parameter can be generalized to be a function of the coordinates, and the laser parameters can also be modulated in time.

Beyond a single ion, the extension to a crystal of many ions whose motions are linearized about their (periodically driven) equilibrium positions would be immediate using analytic expressions for a coupled Mathieu oscillators system \cite{rfions,rfmodes}, with applications ranging from the cooling of 1D chains of ions \cite{PhysRevA.64.063407, morigi2003ion,morigi2001twospecies, fogarty2016optomechanical,kamsap2017experimental, PhysRevLett.119.043001,1367-2630-13-4-043019, PhysRevX.8.021028}, to planar, 2D and 3D crystals in Paul and also Penning traps \cite{Mitchell13111998,Drewsen_Long_Range_Order,
SchilerProteinsPRL,szymanski2012large,tabor2012suitability, mavadia2013control,bohnet2016quantum}, and  applications in quantum information processing \cite{PhysRevLett.81.3631,berkeland1998minimization,PhysRevLett.101.260504,rfmodes, rfions,zigzagexperiment, Landa2014, PhysRevA.90.022332,wang2015quantum,arnold2015prospects, keller2015precise,yan2016exploring, mielenz2016arrays, bruzewicz2016scalable,mihalcea2017study, keller2017optical,keller2018controlling, welzel2018spin,delehaye2018single,PhysRevA.97.062325}. 
The extension of the theory to account for more than two electronic levels could be relevant for different types of ions \cite{PhysRevA.96.012519,janacek2018effect}.

In this work we have focused on adiabatic noise heating, typically applicable to electric field fluctuations in vacuum-operated traps. Collisions of background gas molecules with atomic ions typically induce a nonadiabatic energy change \cite{devoe2009power,rouse2017superstatistical, rouse2018energy}, and their separation in time is much larger than the cooling timescale.
 However, recently objects ranging from large biomolecules, through graphene nanoplatelets to micrometer- and nanometer-scale spheres and diamonds, are being trapped \cite{offenberg2008translational,PhysRevLett.114.123602,PhysRevA.94.010104, doi:10.1063/1.4965859, PhysRevB.82.115441, mihalcea2016multipole, nagornykh2015cooling,PhysRevLett.117.173602, alda2016trapping,aranas2017thermometry, partner2018printed, delord2016electron,delord2017diamonds,
delord2017strong}, and whose dynamics, depending on the pressure in the experiment, can be modelled as Brownian motion. The extension to nonisotropic noise \cite{PhysRevA.92.013414} is immediate. 
Beyond the noise that is inherent to the trap \cite{turchette2000, AnomalousHeatingRate,PhysRevB.89.245435,brownnutt2015ion, ivanov2016high,lakhmanskiy2018observation}, it is also possible to introduce forces with differently tailored noise spectra \cite{PhysRevA.97.020302,sedlacek2018method} and study the ion's dynamics or its probability distribution. Such questions stand at the heart of nonequilibrium formulations of reaction-rate theory (Kramer's escape problem \cite{hanggi1990reaction}) and stochastic resonances \cite{gammaitoni1998stochastic}. 
In combination with laser cooling, a single ion may be captured in a complicated motion \cite{Kaplan09,PhysRevA.82.061402}, and nonequilibrium models of interacting particles coupled to different baths \cite{dotsenko2013two,grosberg2015nonequilibrium,weber2016binary, mancois2018two} could be tested with trapped ions \cite{KinkTickle}, along with various ideas of stochastic, nonequilibrium and active systems.


\begin{acknowledgments}
We thank Vincent Roberdel for very useful discussions. H.L. thanks Giovanna Morigi, Alex Retzker, Shamik Gupta, Anupam Kundu, Oren Raz and Roni Geffen for fruitful discussions.
H.L. acknowledges support by a Marie Curie Intra European Fellowship within the 7th European Community Framework Programme, by IRS-IQUPS of Universit\'{e} Paris-Saclay, and by LabEx PALM under grant number ANR-10-LABX-0039-PALM.

\end{acknowledgments}

\appendix

\section{Transformation to nondimensional variables and trap parameters}\label{Sec:Nondim}

For the numerical calculations presented in this work we use nondimensional units, obtained by rescaling the time $t$ by half the micromotion frequency, $\Omega/2$, and measuring distances using a natural lengthscale of the problem, $w$, which, for our case, is the electrode width in a five-wire surface-electrode Paul trap \cite{rfchaos}). 
The rescaling introduced in \eq{Eq:rescaling} is,
\begin{equation}
 z \to z/w, \qquad t \to \Omega t /2,\qquad  v_z \to v_z/(w\Omega/2).
\label{Eq:rescaling2}
\end{equation}
This rescaling  allows us also to use the ion mass $m$ in order to define a nondimensional momentum, and its charge $e$ to define a nondimensional potential energy $V$ that depends on  an electrostatic voltage $U$,
\be p_z\to p_z/(mw\Omega/2),\qquad U \to U / [  mw^2\Omega^2/(4e)].\ee
The full potential of the trap can be composed of a sum of a few similar potential terms. 
The laser  parameters are similarly rescaled
\be
p_{\rm{r}}\to\frac{{p}_{\rm{r}}}{mw\Omega/2},\quad \Gamma\to\frac{{\Gamma}}{\Omega/2},\quad \Delta\to\frac{{\Delta}}{\Omega/2},
\quad k\to{{k}}{w},\ee
in addition to $\omega_{\rm L} \to{\omega}_{\rm L}/(\Omega/2)$ and ${\Omega}_{\rm{R}} \to{\Omega}_{\rm{R}}/(\Omega/2)$,
together with the rescaled Planck's constant,
\be \hbar \to{\hbar} /(m w^2\Omega/2).\label{Eq:hbar} \ee

For the numerical calculations, we assume throughout this work a  $^9{\rm Be}^+$ ion, and the parameters of \eq{Eq:rescaling2} are
\be w=50\,{\rm \mu m}\qquad \Omega=2\pi\times 100\,{\rm MHz}.\label{eq:physicalparams}\ee
 The other trap parameters are defined by
\begin{equation}
a_z=\frac{4  e U_{\rm{DC}}/{C_z}}{m  w^2 \Omega^2}, \qquad q_5=\frac{2eU_{\rm{rf}}}{m w^2 \Omega^2},
\label{aq5def}
\end{equation}
with $U_{\rm{DC}}$ a voltage on electrodes providing confinement along the trap's symmetry axis, $C_z$ is a nondimensional parameter that characterizes the geometric properties of this harmonic potential, and $U_{\rm rf}$ is the voltage on the rf electrodes. The nondimensional parameter values that we choose are
\be q_5\approx 0.43,\qquad a_z=-0.0002,\qquad q_z\approx 0.16\label{Eq:TrapParamsFP1},\ee
which correspond to $U_{\rm rf}=20\, {\rm V}$,
and this gives 
\be  \nu_z\approx 0.112,\quad \omega_z\approx 2\pi\times 5.60\,{\rm MHz},\label{Eq:TrapParamsFP2}\ee 
with $\omega_z$ the dimensional secular frequency.
For the $^9$Be$^+$ ion, the laser parameters in dimensional units are
 \be \tilde{k}\approx 2\pi / 313\,{\rm nm}^{-1},\qquad \tilde{\Gamma}\approx 120\times 10^6\,{\rm s^{-1}}\label{Eq:LaserParams1},\ee
and we take $\vec{k}=k\hat{z}$, with a transverse laser polarization, giving a spontaneous emission coefficient [using \eq{Eq:mu}],
\be  \mu\equiv \mu_{zz}=2/5,\ee
(where $\mu$ is often denoted by $\alpha$ or $\xi$ in the literature). 
The nondimensional diffusion coefficient $D$ [after the rescaling of \eq{Eq:rescaling}] is given by
\be D=\frac{8 \tilde D}{m^2 w^2 \Omega ^3},\ee
where $\tilde{D}/m$ is the dimensional diffusion coefficient with units of energy increase rate (energy per unit time). Typical measured values are reported as the heating rate $\dot{\tilde{n}}$ in quanta per second (obtained at center of the trap, where the motion can be quantized in terms of a harmonic oscillator).
Hence if the oscillator is heated at a rate in dimensional units of $\dot{\tilde{E}}=\tilde\hbar\omega_z\dot{\tilde{n}}$, it corresponds to the nondimensional diffusion coefficient,
\begin{align}
D=\dot{E}&=\frac{\tilde{\hbar}}{mw^2\pare{{\Omega}/{2}}}\frac{\omega_z}{\pare{\Omega/2}}\frac{\dot{\tilde{n}}}{\pare{\Omega/2}}.\label{Eq:Dfromn}
\end{align}

\section{The fast particle Expansion}\label{Sec:DopplerFastApp}

Expanding the Lorentzian of \eq{Eq:rho_p_z2} in $\Delta$ we get
\be \rho\approx \frac{s}{2}\left[\frac{\Gamma ^2}{4 (\vec{k}\cdot\vec{v})^2+\Gamma ^2}+\frac{8 {\vec{k}\cdot\vec{v}\,} \Gamma ^2 \Delta }{\
\left(4 (\vec{k}\cdot\vec{v})^2+\Gamma ^2\right)^2}
\right],\ee
where this expansion is valid for
\be\left| \vec{k}\cdot\vec{v} \right| \gg \Gamma/2,\qquad \left| \vec{k}\cdot\vec{v} \right| \gg 2 \left|\Delta\right|. \label{Eq:DeltaFast1}\ee
Under the conditions described in \seq{Sec:DopplerFast} (that the laser propagates along  $\vec{k}=k\hat{z}$, and if in addition the $z$ motion is described by the harmonic oscillator of \eq{Eq:hotransf}), the conditions in \eq{Eq:DeltaFast1} are equivalent to \eq{Eq:DeltaFast2}, and
we can use \eq{Eq:hotransf} to perform the integrals in \eqs{Eq:PiIc1D}-\eqref{Eq:PiIIc1D}, to get the harmonic oscillator fast particle limit,
\begin{align} & \Pi_{I}^{\rm f.p.}\approx \hbar \frac{\Delta}{\nu_z}\frac{ s \Gamma^2/4}{k\sqrt{2I\nu_z}} + \hbar^2 k\frac{s\Gamma^2/4}{2\nu_z \sqrt{2I\nu_z}}(1+\mu),\label{Eq:PiIfho} \\&
 \Pi_{II}^{\rm f.p.}\approx \hbar^2\frac{2 (s\Gamma^2/4)}{4\nu_z^2+ \Gamma^2}\left( \frac{\Gamma^3(1+\mu)+4\nu_z^2 \Gamma}{4 \nu_z^2} +  \mu k\sqrt{2I\nu_z}\right)\label{Eq:PiIIfho}.
\end{align}

The first term in \eq{Eq:PiIfho} for $\Pi_I^{\rm f.p.}$ gives cooling (for $\Delta<0$), and coincides with the result derived for the fast particle limit, using quantum harmonic oscillator wavefunctions, in \cite{javanainen1981laser} (where the second term has been neglected). 
The second term in  \eq{Eq:PiIfho} is positive, heating-like, and may counteract the cooling in this asymptotic region of harmonic oscillator motion.
Both terms scale with $1/\sqrt{I}$, and the ratio of the second term to the first term of the drift coefficient $\Pi_I^{\rm f.p.}$, equals $(1+\mu)k p_{\rm r}/(2\Delta)$, and for typical parameters (with $|\Delta|\gtrsim \Gamma/2$) this ratio is small by \eq{Eq:FPHeavyParameter}, implying that the positive drift term can be neglected (but not, however, too close to resonance).

The value of $\Pi_{II}^{\rm f.p.}$ is again the sum of two terms, but these have different asymptotics; the first is constant while the second term scales with $\sqrt{I}$ and gives the dominant functional dependence in the fast particle limit. The latter has the same scaling as that derived in \cite{javanainen1981laser} using the quantum harmonic oscillator wavefunctions, however with a different prefactor, consistent with our model of a classical ion \footnote{To relate the notation of \cite{javanainen1981laser} to our variables, $\mu=1/3$, $N=k\sqrt{2I\nu}/\nu$, and $2\kappa^2=s\Gamma^2/4$. Our result for $\Pi_{II}^{\rm f.p.}$ is smaller by the factor $4\nu_z^2/(\Gamma^2+4\nu_z^2)$. In \cite{javanainen1981laser} it is argued that due to the form of the quantum wavefunctions of the harmonic oscillator, the ion's emission peaks at maximum momentum. This argument does not apply to our model, obtained by assuming a localized ion, oscillating in the trap with well-defined phase-space coordinates.}.
We note for comparison, that the zero lifetime treatment results in the same cooling rate in terms of the energy, but gives a  constant diffusion term instead of a term $\propto \sqrt{I}$, which is an {\it incorrect} result, 
\be \Pi_{II}^{\rm z}\to \hbar^2 \frac{2(s\Gamma^2/4)\Gamma(1+\mu)}{4\nu_z^2},\qquad{(\rm h.o.)}.\label{Eq:PiIIhho}\ee
This diffusion coefficient can be obtained in the limit of both $\Gamma\gg\nu_z$ and $I\to 0$ from \eq{Eq:PiIIfho}, which gives a simple consistency check of the fast particle limit.

\section{Derivatives of the canonical transformation} \label{App:CanonicalDerivatives} 

Using the pseudopotential Hamiltonian,
\be H_{\rm ps}(\zeta,\pi_\zeta)=\frac{1}{2}\pi_\zeta ^2 + V_{\rm ps}(\zeta),\ee and the fact that the Hamiltonian becomes angle-independent in the action-angle coordinates, we can write 
\be \frac{\partial H_{\rm ps}(\zeta,\pi_\zeta)}{\partial \pi_\zeta}=\frac{\partial H_{\rm ps}(I,\theta)}{\partial I}\frac{\partial \Lambda_{\rm ps}(\zeta,\pi_\zeta)}{\partial \pi_\zeta}.\ee 
By rearranging and using the definition of the frequency $\nu(I)=\partial H/\partial I$, we get
\be \frac{\partial \Lambda_{\rm ps}(\zeta,\pi_\zeta)}{\partial \pi_\zeta}= \frac{\pi_\zeta}{\nu(I)}.\label{Eq:derv1}\ee
Similarly,
\be \frac{\partial( \pi_\zeta/\nu)}{\partial \pi_\zeta} = \frac{\partial^2 \Lambda_{\rm ps}(\zeta,\pi_\zeta)}{\partial \pi_\zeta^2}=\frac{1}{\nu}+ \pi_\zeta \frac{\partial (1/\nu)}{\partial \pi_\zeta},\label{Eq:derv2}\ee
and again,
\be \frac{\partial (1/\nu)}{\partial \pi_\zeta}=\frac{\partial (1/\nu)}{\partial I}\frac{\partial \Lambda_{\rm ps}}{\partial \pi_\zeta}=-\frac{1}{\nu^2}\frac{d \nu}{d I}\frac{\partial \Lambda_{\rm ps}}{\partial \pi_\zeta},\ee
which together with \eqss{Eq:derv1}{Eq:derv2} allows to obtain \eq{Eq:TransPseudoPartial} by using \eq{Eq:TransPseudo} to substitute $\partial / \partial p_z \approx \partial / \partial \pi_\zeta$. 

\begin{widetext}

\section{Transformations of the Fokker-Planck equation}\label{Sec:generalFP}

We consider a general from of the Fokker-Planck equation for the distribution $\rho(R,P,t)$, written here for 1D case for simplicity, with canonical coordinates $\{R,P\}$,
\begin{equation}\frac{\partial \rho(R,P,t)}{\partial t}=\mathcal{L}_0(R,P,t) \rho-\sum_{i\in \{R,P\}}\frac{\partial }{\partial i}A_i(R,P,t)\rho+\frac{1}{2}\sum_{i,j\in \{R,P\}}\frac{\partial^2 }{\partial i \partial j}B_{ij}(R,P,t)\rho,\label{generalFP}\end{equation}
where $\mathcal{L}_0(R,P,t)$ is the Liouvillian.
For a  {\it canonical} transformation $\{r,p\}=\{\phi_r\pare{R,P,t},\phi_p\pare{R,P,t}\}$, \eq{generalFP} transforms to \begin{equation}\frac{\partial \rho(r,p,t)}{\partial t}=\mathcal{L}_0(r,p,t)\rho -\sum_{k\in \{r,p\}}\frac{\partial }{\partial k}\tilde{A}_k(r,p,t)\rho+\frac{1}{2}\sum_{k,l\in \{r,p\}}\frac{\partial^2 }{\partial k \partial l}\tilde{B}_{kl}(r,p,t)\rho,\label{App:transformedFP}\end{equation}
 where the coefficients $\tilde{A}_{k}$ and $\tilde{B}_{kl}$ with $k,l\in \{r,p\}$ are given by \footnote{See page 286 of \cite{kampen2007}.}
\begin{equation}\tilde{A}_{k}=\sum_{i\in \{R,P\}} A_i\frac{\partial \phi_k}{\partial i}+\frac{1}{2}\sum_{i,j\in \{R,P\}} B_{ij}\frac{\partial^2 \phi_k}{\partial i \partial j},\qquad\tilde{B}_{kl}=\sum_{i,j\in \{R,P\}}B_{ij}\frac{\partial \phi_k }{ \partial i}\frac{\partial \phi_l }{\partial j},  \label{App:transfoB}\end{equation}
and we note that the Liouvillian in the new coordinates has to be constructed using the Hamiltonian in the transformed coordinates, $K(r,p,t)=H(r,p,t)+\partial F/\partial t$ with $F$ the generating function of the canonical transformation. More general, time-dependent but noncanonical transformations, and an averaging treatment, are presented in \cite{PhysRevE.67.061111}.


\end{widetext}

\bibliographystyle{hunsrt}
\bibliography{rf_cooling}

\end{document}